\documentclass[fleqn,usenatbib]{mnras}

\usepackage{newtxtext,newtxmath}
\usepackage{multirow} %used for tables
\usepackage{booktabs} %used for hlines in tables
\usepackage{longtable}
\usepackage{soul} %for highlighting text during colaborative review
\usepackage[T1]{fontenc}

\DeclareRobustCommand{\VAN}[3]{#2}
\let\VANthebibliography\thebibliography
\def\thebibliography{\DeclareRobustCommand{\VAN}[3]{##3}\VANthebibliography}

\usepackage{graphicx}	% Including figure files
\usepackage{amsmath}	% Advanced maths commands
\title[The magnetic and RV variability of V889 Her]{The variable magnetic field of V889 Her and the challenge of detecting exoplanets around young Suns using Gaussian process regression}

\author[E. L. Brown et al.]{
E. L. Brown,$^{1,2}$\thanks{E-mail: emma.brown@usq.edu.au}
S. C. Marsden,$^1$
S. V. Jeffers,$^3$
A. Heitzmann,$^{1,4}$
J. R. Barnes$^{5}$ and
C. P. Folsom$^{6}$
\\
$^{1}$Centre for Astrophysics, University of Southern Queensland, Toowoomba, QLD, 4350, Australia\\
$^{2}$School of Access Education, Central Queensland University, Cairns, QLD, 4870, Australia\\
$^{3}$Max Planck Institut for Solar System Research, Justus-von-Liebig-Weg 3, 37077 G\"ottingen, Germany\\
$^{4}$Observatoire Astronomique de l’Université de Genève, Chemin Pegasi 51, 1290 Versoix, Switzerland \\
$^5$Department of Physical Sciences, The Open University, Walton Hall, Milton Keynes MK7 6AA, UK\\
$^6$Tartu Observatory, University of Tartu, Observatooriumi 1, Tõravere 61602, Estonia}

\date{Accepted XXX. Received YYY; in original form ZZZ}

\pubyear{2024}

\begin{document}
\label{firstpage}
\pagerange{\pageref{firstpage}--\pageref{lastpage}}
\maketitle

% Abstract of the paper
\begin{abstract}
Discovering exoplanets orbiting young Suns can provide insight into the formation and early evolution of our own solar system, but the extreme magnetic activity of young stars obfuscates exoplanet detection. Here we monitor the long-term magnetic field and chromospheric activity variability of the young solar analogue V889\,Her, model the activity-induced radial velocity variations and evaluate the impacts of extreme magnetism on exoplanet detection thresholds.  We map the magnetic field and surface brightness for 14 epochs between 2004 and 2019. Our results show potential 3-4\,yr variations of the magnetic field which evolves from weak and simple during chromospheric activity minima to strong and complex during activity maxima but without any polarity reversals. A persistent, temporally-varying polar spot coexists with weaker, short-lived lower-latitude spots. Due to their different decay time-scales, significant differential rotation and the limited temporal coverage of our legacy data, we were unable to reliably model the activity-induced radial velocity using Gaussian Process regression. Doppler Imaging can be a useful method for modelling the magnetic activity jitter of extremely active stars using data with large phase gaps.  Given our data and using Doppler Imaging to filter activity jitter, we estimate that we could detect Jupiter-mass planets with orbital periods of $\sim$3\,d. A longer baseline of continuous observations is the best observing strategy for the detection of exoplanets orbiting highly active stars. 
\end{abstract}

% Select between one and six entries from the list of approved keywords.
% Don't make up new ones.

%%%%%%%%%%%%%%%%%%%%%%%%%%%%%%%%%%%%%%%%%%%%%%%%%%

%%%%%%%%%%%%%%%%% BODY OF PAPER %%%%%%%%%%%%%%%%%%
\section{Introduction}

Young solar analogues show extreme magnetic activity that contrasts against the solar case, yet our Sun is thought to have behaved similarly during its infancy. By studying the magnetism of young solar analogues we can look back at the Sun's possible early main sequence life and see how the conditions would have been within our own space environment.  The detection of exoplanets around these stars could also give insight into the formation and early evolution of planetary systems around stars like our own, making young solar analogues important targets for both magnetic activity studies and in the search for exoplanets.

Today's Sun undergoes $\sim22$\,yr magnetic cycles, wherein the large-scale magnetic field polarity reverses twice (every $\sim11$\,yr) and between polarity reversals the magnetic field is converted from a dipolar, poloidal structure at magnetic minima, to a slightly more toroidal and complex geometry at magnetic maxima \citep{DeRosa2012}. The emergence of magnetic flux at the solar surface manifests in the growth and decay of dark spots, bright faculae and plages throughout an 11\,yr `spot cycle', and also gives rise to a chromospheric activity cycle that is measurable using the emissions in the cores of the chromospheric \ion{Ca}{ii} H and K spectral lines \citep{Wilson1978}. Solar-like magnetic cycles have been identified in three other mature Sun-like stars based on periodic reversals of the large-scale magnetic field polarity that coincide with chromospheric activity cycles; 61 Cyg A (1.3 - 6 Gyr, K5V, $\sim14$\,yr magnetic cycle, \citealt{BoroSaikia2016,BoroSaikia2018a}), $\tau$\,Boo (0.9 Gyr, F8V, $\sim240$\,d cycle, \citealt{Fares2009,Mengel2016,Jeffers2018}) and HD 75332 (1.88 Gyr, F7V, $\sim1.06$\,yr cycle, \citealt{Brown2021}).  

Young solar-type stars have shown comparatively chaotic magnetic behaviour, with magnetic fields that are an order of magnitude or more stronger than those seen in mature stars, often multiple coexisting activity cycles \citep{Olah2016,BoroSaikia2022,Jeffers2022} and polarity reversals that do not necessarily correspond to chromospheric activity cycles \citep{BoroSaikia2022}. One of the closest young solar-type stars, the $\sim0.5$\,Gyr old \citep{Barnes2007} K2 dwarf $\epsilon$ Eridani, potentially alternates between chaotic and cyclic variability \citep{Metcalfe2013,Petit2021}, and the results from \citet{Jeffers2022} are consistent with a dynamo of two cycles where the shorter 3\,yr cycle is a modulation of the longer 13\,yr cycle. Young solar-type stars typically show strongly toroidal and axisymmetric magnetic fields \citep{See2016,Folsom2018a} compared to the Sun's dominantly poloidal magnetic field, with the most rapidly rotating main-sequence stars often showing strong rings of azimuthal field around the rotational axis (e.g. \citealt{Marsden2006,Jeffers2008,Jeffers2011}). Tomographic imaging of surface temperature contrasts, known as Doppler Imaging (DI), reveals large polar spots, or collections of spots, on young stars that are sustained for many years, in addition to comparatively short-lived low- to mid-latitude spots (e.g. \citealt{Marsden2006,Jeffers2008,Jeffers2011,Lehtinen2022}). This is in stark contrast to the Sun, where spots are restricted to low latitudes, cover only 0.3 to 0.4 percent of the solar surface at any time \citep{Solanki2004} and last only days to months.

The extreme magnetic activity of young stars impacts on the ability to detect exoplanets using the radial velocity (RV) method, which is an important adjunct to other exoplanet detection techniques because it places unique constraints on the planetary mass. Exoplanet detection using the RV method relies on identifying Doppler shifts in stellar spectral lines to measure small changes in the RV of the star that are induced by orbiting objects. Magnetic activity complicates the detection of true Doppler shifts because spots and bright regions cause asymmetries in spectral line profiles that shift the line centre as they rotate in and out of view. These line shifts result in apparent, quasi-periodic RV variations with amplitudes up to hundreds of m\,s$^{-1}$, which can obscure or even mimic the RV signatures of exoplanets \citep{Nava2020,Barnes2011,Jeffers2014b}. Other magnetic activity signatures, including a net convective blueshift across the stellar disk \citep{Meunier2010} and unresolved magnetic spots \citep{Lisogorskyi2020}, can each impart RV amplitudes as large as 10 m\,s$^{-1}$, with this alone dwarfing the stellar RV signatures of planets up to 20\,$\rm{M_{\oplus}}$.  

\begin{figure}
    \centering
    \includegraphics[width=\linewidth]{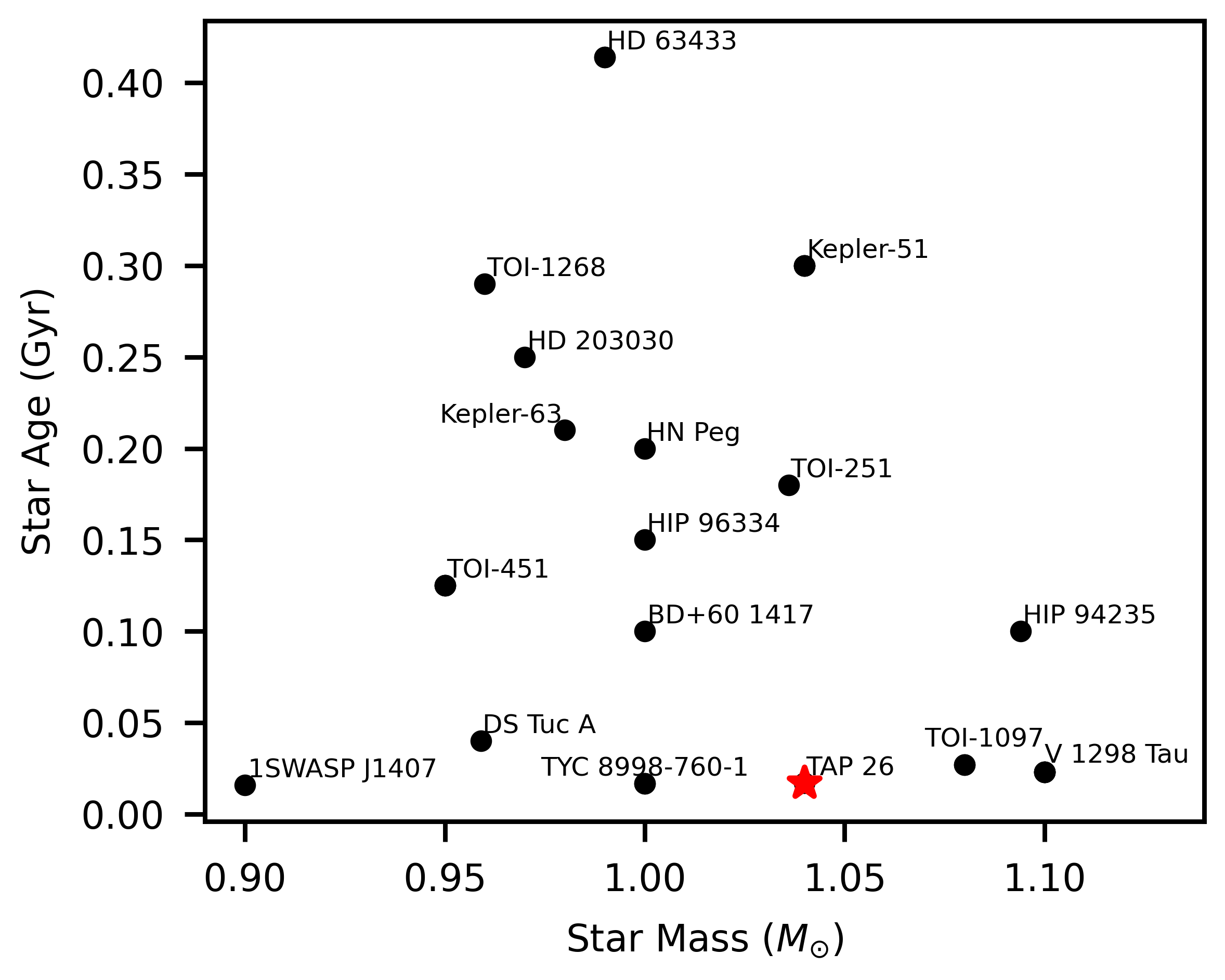}
    \caption{Age versus mass for planet-hosting stars with masses within 10 percent of $\rm{M_{\odot}}$ and ages below $\sim500$\,Myr. Only the star TAP26 (red) has had an exoplanet detected using only stellar RVs. For the remaining stars (black) exoplanets were detected using transits, direct imaging or astrometry. Data taken from The Extrasolar Planets Encyclopaedia\protect\footnotemark.}
    \label{fig:RV_detections}
\end{figure}

Young solar analogues make particularly difficult targets in the search for exoplanets, being near-solar mass which makes true exoplanet RV signatures small, and highly active, making activity-induced RV amplitudes large. However, \citet{Jeffers2014b} predicts that hot Jupiters (HJs) can be detected for young G and K dwarfs if they are observed over a sufficient time-base of 100 epochs. So far, for stars with masses within 10 per cent of $M_{\odot}$ and ages up to 500\,Myr, shown in Figure \ref{fig:RV_detections}, only one has had an exoplanet detected using the RV method alone. 

\footnotetext{\url{http://exoplanet.eu/}}

Methods to model and filter activity-induced RVs, such as the popular method of Gaussian Process Regression (GPR) and the `DI' approach (see Section \ref{sec:Analysis}), have mostly been applied to near-solar mass stars with low-to-moderate levels of magnetic activity (e.g. \citealt{Haywood2014,Heitzmann2021}). GPR has also been tested against simple simulated spot patterns \citep{Perger2021}, but the reality is that young solar analogues have more complex spot topologies, with both long-lived polar spots and rapidly varying lower-latitude features (e.g. \citealt{Marsden2006,Senavc2021,Willamo2022_5SolarTypeStars}). \citet{Heitzmann2021} tested the techniques of DI and GPR on the moderately active, 17-32\,Myr, 1.3\,$M_{\odot}$ G2 dwarf HD 141943, using legacy observations from the non-stabilized University College London Echelle Spectograph (UCLES\footnote{https://www.aao.gov.au/science/instruments/current/UCLES/observing}) at the Anglo Australian Telescope (AAT). They found GPR to be the more successful technique, able to recover synthetic planetary signals with semi-amplitudes $\geq100 $m\,s$^{-1}$ (i.e. a 1\,M$_{\rm{Jup}}$ planet with orbital period of 5\,d, for HD 141943). However, the spot pattern in HD 141943 that consists of low- and high-latitude features with similar strengths and area coverage, still contrasts against the intense polar spots that are ubiquitous among the most active young Suns and their significantly weaker low-latitude features. For active, low-mass stars, such as the 0.344\,$M_{\odot}$, M3.5 dwarf EV Lac \citep{Jeffers2022}, the use of low-resolution DI activity models to remove the spot-induced RV component have been shown to improve the planet-mass sensitivity by a factor of three.  The effectiveness of the DI and GPR methods for modelling the activity-induced RVs of highly active solar-mass stars has not yet been evaluated.  

The young solar analogue V889 Her (HD 171488, G2V, \citealt{Montes2001}, 30-50\,Myr old, $\sim$1.06\,$\rm{M_{\odot}}$, \citealt{Strassmeier2003}) is one of the brightest solar analogues that is rotating fast enough for active regions on its surface to be resolved using the technique of high-resolution DI. This makes it an important target to gain insight into the magnetic dynamos operating inside the most active Sun-like stars, and an ideal candidate to model activity-induced RVs using both the DI and GPR methods. V889 Her has been monitored spectropolarimetrically with the NARVAL\footnote{http://www.ast.obs-mip.fr/projets/narval/v1/} spectropolarimeter at T\'elescope Bernard Lyot (TBL, at the Observatoire Pic du Midi, France, \citealt{Auri2003}), HARPSpol (High Accuracy Radial velocity Planet Searcher polarimeter) at the 3.6-m ESO telescope (La Silla Observatory, Chile, \citealt{HARPSpol}) and the SEMEL polarimeter (SEMpol) at the AAT (Siding Springs, Australia, \citealt{AAT}). This large data set provides an opportunity to study the long term variability of magnetism in V889 Her, and evaluate the impact of an extended baseline of observations on exoplanet detection capabilities for young solar analogues.  

The large-scale magnetic field structure and brightness topology of V889 Her have been reconstructed in a number of previous studies \citep{Strassmeier2003,Marsden2006,Jeffers2008,Jarvinen2008,Huber2009,Frasca2010,Jeffers2011,Willamo2022}, which have shown a strongly toroidal field geometry and a large spot (or collection of small, unresolved spots) that covers the visible pole of the star. The polar spot seems to have persisted from at least 1998 to 2011 \citep{Strassmeier2003,Willamo2022}. %The previous studies give conflicting measures of differential rotation (DR), with \citet{Marsden2006}, \citet{Jeffers2008} measuring high levels of DR.
Long term photometric monitoring by \citet{Lehtinen2016} and \citet{Willamo2019} has shown V889 Her to have a brightness cycle with a period of $\sim7-9$\,yr, but long-term S-index observations have not been available to detect a chromospheric activity cycle. \citet{Willamo2019} did not find any consistent anti-correlation between the long-term brightness variations and spot filling factor (percent of the stellar surface covered in spots), as would be expected for a rapidly rotating star \citep{Radick1990}, which they attributed to unreliable filling factor estimates. \citet{Willamo2019} also found that V889 Her could have experienced a grand maximum in its magnetic activity between 2007 and 2017, based on a long-term minimum in brightness (if taken as a maximum in spot coverage/activity). Additionally, V889 Her has been monitored with NASA/MIT’s Transiting Exoplanet Survey Satellite (TESS\footnote{https://tess.mit.edu/}). Potential transit signals have been flagged in the TESS data validation report summaries, with periods ranging between $\sim10$ and 382\,d, but none of these candidates have been elevated to a TESS Object of Interest (TOI). 

In this work we study the magnetic and chromospheric activity variability of V889\,Her, and model the activity-induced stellar RV shifts in an attempt to uncover the true stellar RV and search for RV shifts induced by exoplanets. The remainder of this paper is organised as follows; in Section \ref{sec:Data} we describe the data used in this study. In Section \ref{sec:Analysis} we outline the techniques used to reconstruct the large-scale magnetic field and brightness topology of V889 Her, extract precise RVs, chromospheric \ion{Ca}{ii} H\&K activity and model the activity jitter. In Section \ref{sec:results_discussion1} we present the results of brightness and magnetic field modelling, and analyse the long-term evolution of the large-scale magnetic field and brightness topology of V889 Her. The activity-induced RV variability is discussed in Section \ref{sec:results_discussion2}, as well as the challenges encountered in modelling the activity-induced RVs using GPR and DI to search for exoplanets. Finally, our conclusions are summarized in Section \ref{sec:conclusions}. 
\section{Data}\label{sec:Data}

\subsection{Spectropolarimetry with NARVAL, HARPSpol and SEMpol}

Our data include both previously published and new observations of V889\,Her covering the period from 2004 May to 2019 July. We used observations from the NARVAL spectropolarimeter, HARPSpol polarimeter in conjunction with the HARPS spectrograph,  and SEMEL polarimeter coupled with UCLES. HARPS spectra have a resolution of $\sim$110000 and spectral coverage from 3800 to 6900 {\AA} compared to the data from NARVAL with a resolution of $\sim$65000 and spectral coverage from 3700 to 11000 {\AA}, and UCLES spectra with a resolution of $\sim$70000 and wavelength coverage 4400 to 6800 {\AA}. The observations are summarized in Table \ref{tab:observation_summary} and further details are provided in the supplementary materials, with previously published data indicated where applicable. 

Each of the instruments is configured for spectropolarimetry with the polarimeter mounted at the Cassegrain focus, to split incoming circularly polarized (Stokes {\it{V}}) light into two orthogonally-polarized beams, both of which are fibre-fed to the spectrograph and recorded on the detector. A single Stokes {\it{V}} observation consists of a series of 2 or 4 sub-exposures, where the positions of the orthogonally-polarized beams are switched between sub-exposures. The Stokes {\it{V}} observation is determined by dividing sub-exposures with orthogonal polarization states, which removes spurious polarization signals to a first-order degree. Adding sub-exposures provides the unpolarized, Stokes {\it{I}} observation. For series' of 4 sub-exposures, a `Null' spectrum can also obtained by dividing spectra with identical polarization states, and this provides a measure of noise and a diagnosis of the reliability of the polarimetric measurement. For further information on this process see \citet{Donati1997}.

\subsection{Data reduction and calibration}\label{sec:data_reduction_and_calibration}
The observations were reduced and calibrated using the automated software package {\sc{libre-ESpRIT}} (based on {\sc{ESpRIT}}, Echelle Spectra Reduction: an Interactive Tool, \citealt{Donati1997}). {\sc{libre-ESpRIT}} reduces the observations using  bias and flat-field exposures averaged from series' of exposures taken on the observing night. A first-order wavelength calibration is carried out using a Thorium-Argon arc lamp exposure. For NARVAL and UCLES spectra, the wavelength calibration is then further refined using telluric lines as RV references, as was first done by \citet{Donati2003_telluriclines}. Telluric corrections are carried out for each observation individually, and this reduces the relative RV shifts to 100 m\,s$^{-1}$ for UCLES spectra \citep{Donati2003_telluriclines} and 30 m\,s$^{-1}$ for NARVAL spectra \citep{Donati2008, Moutou2007}.  For HARPS spectra we adopt an instrumental precision of 5\,m\,s$^{-1}$ based on \citet{Trifonov2020}.%\citep{Fischer2016}.}%, UCLES spectra have an accuracy of $3\rm{ms^{-1}}$ \citep{Fischer2014} and NARVAL spectra $\sim30\rm{ms^{-1}}$ \citep{Moutou2007}.

\subsection{LSD spectral line profiles}\label{sec:LSD}
We used the technique of Least-squares deconvolution (LSD, \citealt{Donati1997}) to derive high signal-to-noise spectral line profiles from the observed spectra. LSD assumes that the spectral line asymmetries caused by magnetic activity are roughly equal across all spectral lines. Therefore, the entire observed spectrum of an active star can be thought of as a convolution of a `mean' spectral line profile (i.e. the LSD profile) and the spectrum of the star when it is magnetically inactive. To model the spectrum of V889 Her we used a line mask for a $T_{\mathrm{eff}}=5750\,K$, $\log g=4.50$ and $\log\textrm{(M/H)}=0.00$ star from \citet{marsden2014}. The line mask excluded lines with depth less than 10 per cent of the continuum, and strong lines such as the chromospheric \ion{Ca}{ii} H\&K lines. LSD was performed using the automated routine included in {\sc{libre-ESpRIT}} \citep{Donati1997}, and using the LSD normalization parameters from \citet{marsden2014}. Uncertainties for each of the LSD spectral pixels are scaled up from the spectral uncertainties as described in \citet{Wade2000,Wade2000b}.

\subsection{Photometry}
V889 Her was observed by TESS in 2-minute short-cadence integrations in sectors 26, 40 and 53. The Pre-search Data Conditioned Standard Aperture Photometry (PDCSAP) light curves are shown in Figure \ref{fig:TESS}, which show relative flux variations of up to $\sim0.15\times10^{5}$\,e\,s$^{-1}$ which are assumed to be astrophysical. 

V-band photometry obtained with the Tennessee State University T3 0.4\,m Automatic Photoelectric Telescope (APT) located at Fairborn Observatory, Arizona, and spanning the years 1994 to 2018, was also previously published in \citet{Lehtinen2016} and \citet{Willamo2019}, and provided on {\sc{VizieR}} \citep{Vizier:Willamo2019}. The observations were made at a cadence of $\sim1-3$\,d. They were made differentially with respect to the comparison star HD 171286 and the check star HD 170829, which is a constant star that is observed at the same time as the variable target to confirm that the comparison star has constant brightness.   For further information on the acquisition and reduction of the differential photometry see \citet{Lehtinen2016} and \citet{Willamo2019}. 

\section{Analysis}\label{sec:Analysis}

\subsection{Surface brightness and magnetic field mapping}

\subsubsection{Image reconstruction with {\sc{ZDIpy}}}\label{sec:ZDI}

The tomographic imaging technique of DI and its derivative, Zeeman Doppler Imaging (ZDI), reconstruct the brightness contrasts and 3-dimensional large-scale magnetic field across the stellar surface from series' of Stokes {\it{I}} and {\it{V}} LSD profiles respectively. In this work we used the {\sc{ZDIpy}} code by \citet{Folsom2018a}, which is a python implementation of the extensively used C code described in \citet{Donati1997b} and uses a principle of maximum entropy image reconstruction, wherein the feature content of the reconstructed map is minimized. The resulting spot or magnetic map provides a lower limit for the amount of spot coverage or magnetic field strength, and reliably recovers the fractions of different magnetic field components \citep{Lehmann2019}. 

\begin{table*}
\caption{Summary of spectropolarimetric observations of V889 Her used for DI and ZDI. For each epoch, columns list the number of observations, the number of rotational cycles covered by the observations, the HJD of the first (HJD start) and last (HJD end) Stokes {\it{V}} observations, and the HJD corresponding to rotational cycle 0 for the epoch (close to the mid-point of the data, but an integer multiple of rotations from HJD=2453200). The final columns show the instrument that the observations were sourced with and indicate whether the data has been previously published. Full details for individual observations, including observations not used for tomographic imaging, are provided as supplementary materials. * The 2005 May data originally published in \citet{Jeffers2008} were not reanalysed here, but have been included in Table \ref{tab:observation_summary} for completeness.} 
\begin{tabular}{lccccccc}
\toprule
\multirow{2}{*}{Map} & \multirow{2}{*}{Observations} & \multirow{2}{*}{Rotations} & HJD start & HJD end & Mid HJD & \multirow{2}{*}{Instrument} & \multirow{2}{*}{Ref.} \\ 
  & & & (+2453200) & (+2453200) & (+2453200)  & & \\ \midrule
2004 Sept.		&	5	&	3.00    &	72.91 & 76.90 & 	74.37	&		SEMpol & \citet{Marsden2006} \\
2005 May*		&	10	&	7.61    &	322.45 & 332.35 & 	326.50	&		MuSiCoS & \citet{Jeffers2008} \\
2007 May		&	29	&	3.87	& 1042.47	&	1047.61	& 1045.14	&	 NARVAL & \citet{Jeffers2011}\\
2007 Nov.		&	14	&	3.81	& 1213.24	&	1218.31	& 1215.12	&	 NARVAL & \citet{Jeffers2011}\\
2008 May		&	11	&	3.04	& 1413.46	&	1417.50	& 1415.65	&	 NARVAL & \citet{Jeffers2011}\\
2009 May		&	12	&	3.05	& 1779.48	&	1783.53	& 1780.85	&	 NARVAL & This study \\
2010 Aug.		&	8	&	3.09	& 2218.58	&	2222.69	& 2220.42	&	HARPS & This study\\
2011 May 14-16	&	10	&	1.64	& 2496.76	&	2498.93	& 2497.97	&	HARPS & \citet{Willamo2022}\\
2011 May 17-19	&	9	&	1.64	& 2499.74	&	2501.92	& 2500.62	&	HARPS & \citet{Willamo2022}\\
2013 Sept.		&	5	&	3.09	& 3347.49	&	3351.58	& 3349.22	&	HARPS & This study\\
2018 June		&	5	&	3.86	& 5095.43	&	5100.56	& 5098.19	&	 NARVAL & This study\\
2018 July		&	12	&	8.20	& 5117.51	&	5128.40	& 5123.42	&	 NARVAL & This study\\
2019 June		&	12	&	6.75	& 5456.43	&	5465.39	& 5460.74	&	 NARVAL & This study\\
2019 July 10-19	&	10	&	6.78	& 5475.38	&	5484.39	& 5479.33	&	 NARVAL & This study\\
2019 July 20-29	&	13	&	6.76	& 5485.42	&	5494.39	& 5489.95	&	 NARVAL & This study\\
 \bottomrule
\end{tabular}
\label{tab:observation_summary}
\end{table*}

{\sc{ZDIpy}} divides the stellar disk into equal-area tiles.  For each tile the Stokes {\it{I}} profile is modelled with a psuedo-Voigt profile (convolution of a Gaussian and Lorentzian, approximated according to \citealt{Humlicek1982}), which is a more realistic approximation of the Stokes {\it{I}} profile compared to a Gaussian \citep{Folsom2018a}. The Stokes {\it{V}} profile is derived using the weak-field approximation \citep{Donati1997}. The radial, azimuthal and meridional magnetic field components are each represented as a series of spherical harmonics, with the order of the harmonic expansion limited by a factor $l_{max}$ that is related to the complexity of the magnetic field (e.g. $l=1$ for a dipolar field, $l=2$ for quadrupolar field etc.). The full, disk-integrated line models are derived by summing the contributions of all tiles across the stellar surface. The contribution of each tile is scaled by its projected area and relative Doppler shift, taking into account the impact of limb darkening. The brightness or magnetic field (depending on which is being fitted) in each tile is iteratively varied, and the disk-integrated model computed. A model surface brightness or magnetic field map is simultaneously reconstructed from the model line profiles for each iteration, using a principle of maximum entropy image reconstruction, wherein the image feature content is minimized \citep{Donati2006}. The final maximum entropy map is that which achieves a target reduced-$\chi^2$ fit to the observations. We used a target reduced-$\chi^2$ value of 0.1 for Stokes {\it{I}} fitting, and 0.9 for Stokes {\it{V}} fitting (which is just below the noise level).  It is usual to over-fit the Stokes {\it{I}} line profiles  because the signal-to-noise ratio for the LSD Stokes {\it{I}} profiles is underestimated \citep{Petit2004,Wade2000,Wade2000b}. 

ZDIpy is capable of modelling the stellar surface brightness using a ‘filling factor’ or spot-only approach, whereby the star is modelled as a quiet surface (relative brightness = 1) with dark regions (relative brightness < 1), or the surface can be modelled as a combination of bright
(relative brightness > 1) and dark (relative brightness < 1) regions. The spot-only approach has been used in all previous spot studies of V889 Her (\citealt{Strassmeier2003, Marsden2006, Jeffers2008, Jarvinen2008, Huber2009, Frasca2010, Willamo2019, Willamo2022}) so here we use both methods to compare the resulting spot maps and determine if bright features like faculae are negligible in the DI analysis. Brightness mapping is a somewhat simplified approach compared to temperature mapping; in brightness mapping each surface pixel has a relative brightness, while in temperature mapping each pixel has a spot and a quiescent photosphere temperature, as well as a filling factor for the portion of the pixel covered by spots (e.g. \citealt{CollierCameron2001}).  %Sometimes this filling factor is assumed to be unity and spot temperatures are reconstructed, at the risk of underestimating spot temperatures, and sometimes spot and quiescent photosphere temperatures are assumed and filling factors are reconstructed.  
In a single line analysis, this spot temperature and filling factor are effectively degenerate, an issue that brightness mapping sidesteps by using a simplified model with only one free parameter per pixel.  In brightness mapping spots are assumed to only produce changes in line shape, not equivalent width.  Thus the reconstructed brightness maps are essentially only sensitive to variations in the LSD profile shapes, not equivalent widths.  

ZDIpy can model the magnetic field with or without taking into account a brightness map. When a brightness map is used as input for the magnetic field inversion, the code adjusts the inversion to account for the suppression of magnetic flux in dark regions. We tested our magnetic inversions when using spot-only and spot + bright maps as input, and found no significant difference in the recovered magnetic field. The magnetic maps presented in our results take the spot-only maps as input. 

ZDIpy can account for differential rotation of surface features using a solar-like differential rotation law:
\begin{equation}\label{DR_equation}
    \Omega (\theta) = \Omega_{\rm{eq}} - \rm{d}\Omega \sin^2\theta
\end{equation}
where $\Omega (\theta)$ is the angular rotation rate at a latitude $\theta$, $\Omega_{\rm{eq}}$ is the angular rotation rate at the equator, and d$\Omega$ the difference in rotation rate between the equator and the poles (i.e. the differential rotation rate), all measured in $\rm{rad\,d^{-1}}$. 

Table \ref{tab:observation_summary} provides a summary of the spectropolarimetric observations organised into 14 epochs, each with a sufficient number of observations (at least 5) and adequate rotational phase coverage to map the stellar surface using DI and ZDI. The time-spans covered by the data subsets range from 1.64 rotation cycles ($\sim3\,$d) in 2011 May 14-16 and 2011 May 17-19, through to 8.20 rotational cycles ($\sim\rm{11\,d}$) in 2018 July. Even when adopting significant differential rotation in our models, we found that we needed to split the 2011 May data into small sub-sets to achieve a good fit of the ZDI model to the observations, consistent with \citet{Willamo2022} and the rapid surface evolution seen in the TESS light curve for Sector 40 in Figure \ref{fig:TESS}. This suggests that field evolution occurs rapidly, on a time-scale as short as $\sim$3\,d. For other epochs, such as 2018 July through to 2019 July, we were able to easily fit to the noise level with much longer series' of observations. The difference could be due to the higher resolution of the HARPSpol data, which may better recover the evolution of small scale magnetic features, or possibly evolution of the stellar surface is not always as rapid as in 2011 May. We tested shorter series' of observations for 2018 June through to 2019 July, but did not find that this significantly affected the results we report in Section \ref{sec:Longterm_evolution_of_the_magnetic_field_and_brightness_topology}.  The magnetic and bright/dark features in our reconstructions for these epochs (Figures \ref{fig:maps1} to \ref{fig:maps3}) do not show any obvious smearing that would suggest more rapid evolution of the magnetic field was occurring. %Comparatively, when we tested a combined 2011 May data set, covering 6\,d, the reconstructed field was considerably stronger and more complex, with a weaker poloidal component. 

\subsubsection{Stellar and model parameters}

We modelled the Stokes {\it{I}} LSD line profiles using a solar line model with the properties shown in Table \ref{tab:model_parameters}. The line model parameters were fitted to a solar LSD profile, derived from a NARVAL spectrum of sunlight reflected from the moon. The line model is considered to be a suitable approximation for V889 Her which has a similar spectral type to the Sun.

{\sc{ZDIpy}} also requires input of the centre-of-mass RV of the star, its line-of-sight projected rotational velocity ($v\sin i$), inclination angle ($i$), equatorial rotation period ($\rm{P_{eq}}$) and rate of differential rotation ($\rm{d\Omega}$). To select appropriate parameters we used a process of $\chi^2$ minimization, whereby we iteratively varied input parameters and recomputed the stellar model to identify the parameters that minimize the reduced-$\chi^2$ fit to the observations. 

For the RV, $v\sin i$ and inclination angle, we varied the parameters individually and fitted to the Stokes {\it{I}} LSD line profiles. The best-fit RV fluctuated slightly between epochs, as shown in Table \ref{tab:results}. It is not clear if these long-term RV shifts are true centre-of-mass fluctuations, but to optimize the DI and ZDI models we have adopted the best-fit RV on an epoch-to-epoch basis.  The best-fit $v\sin$i and inclination angle are shown in Table \ref{tab:model_parameters}, and are reasonably consistent with previous work \citep{Strassmeier2003,Marsden2006,Jeffers2008, Jeffers2011,Willamo2019}. The inclination angle is typically the most difficult parameter to constrain using $\chi ^2$ minimization. The $60\pm5\degr$ we derived using $\chi ^2$ minimization is consistent within the error bars with an inclination angle of 64.5$\degr$ derived using our measured equatorial rotation period and $v\sin i$, adopting a stellar radius of 1.09$R_{\odot}$ \citep{Strassmeier2003} and by assuming solid-body rotation. As discussed by \citet{Willamo2022}, the possible detection of transiting exoplanets by TESS suggests a higher inclination angle if we assume that the planet's orbital axis is aligned with the star's axis of rotation. We tested our magnetic field and surface brightness reconstructions using a greater inclination angle of $85\degr$ and the models converged at much higher reduced-$\chi^2$ values, indicating a poor fit of the model to the data. Given that none of the potential transits detected by TESS have been elevated to a TOI, we have adopted the best-fit inclination angle of $60\degr$ for ZDI.

To determine the optimal equatorial rotation period and rate of differential rotation we varied the parameters together according to Equation \ref{DR_equation}. We fit this using the Stokes {\it{V}} profiles, since V889 Her shows more latitudinal distribution of magnetic features compared to spots, which are mostly observed over the pole both in our study and in previous work \citep{Strassmeier2003,Marsden2006,Jeffers2008, Jeffers2011,Willamo2019,Willamo2022}. The HJD corresponding to rotational cycle zero (i.e. the date where the magnetic field is assumed to be unsheared by differential rotation) is given in Table \ref{tab:observation_summary} for each epoch. 

\begin{figure*}
    \centering
    \includegraphics{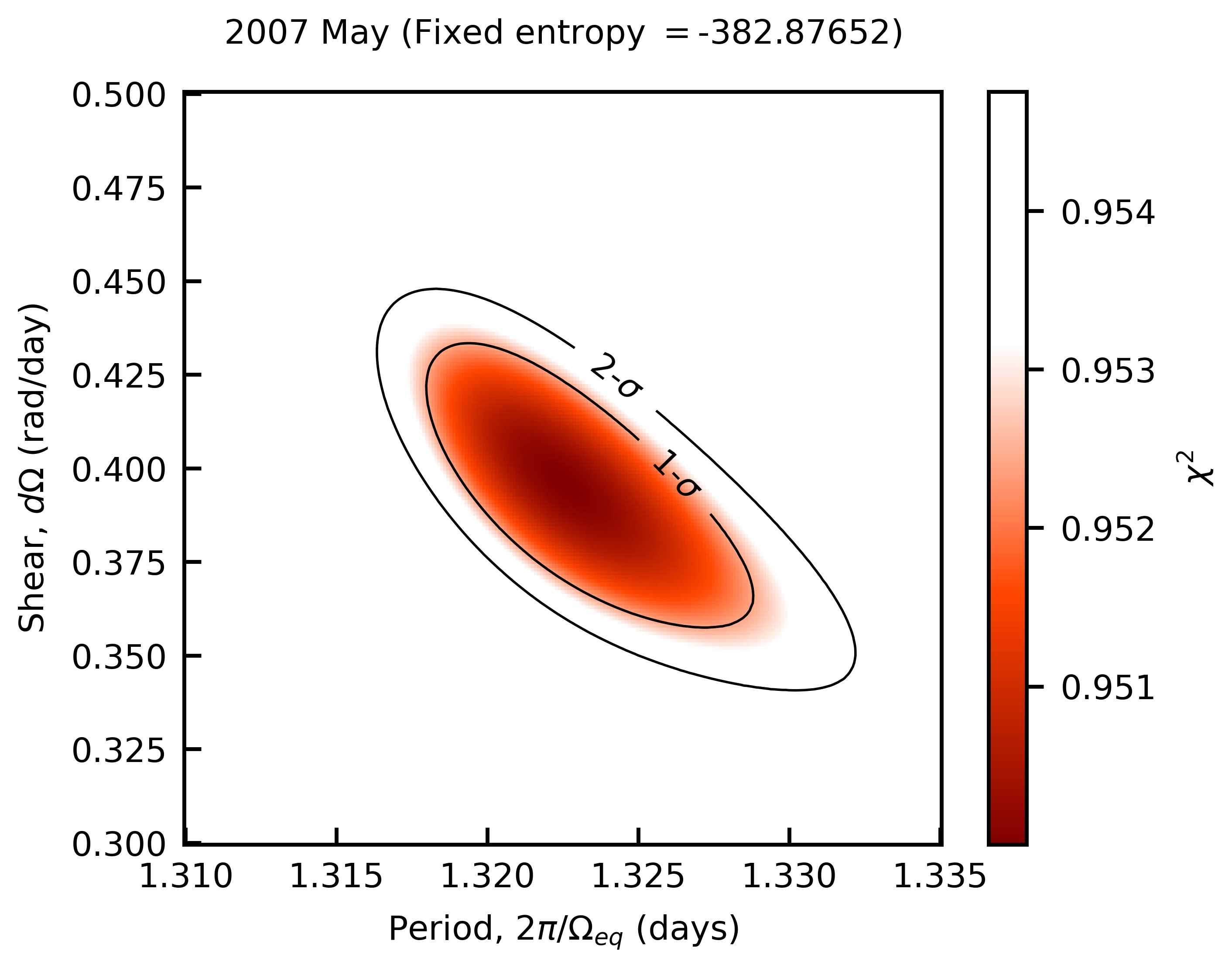}\includegraphics{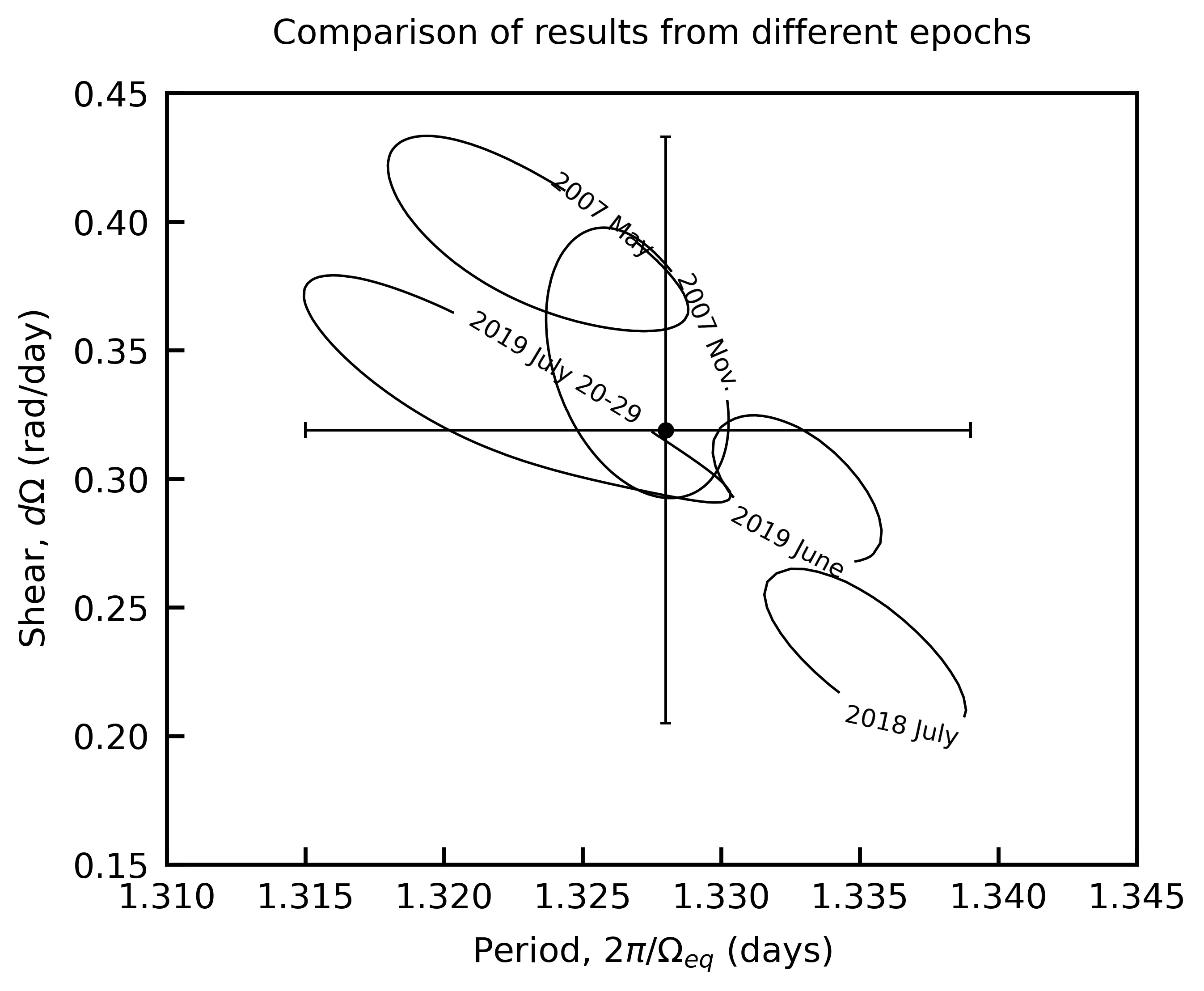}
    \caption{Left: Example plot of the reduced-$\chi^2$ landscape when fitting to the Stokes {\it{V}} profiles from 2007 May. The darkest red region represents the best-fit combination of the equatorial rotation period (d) and rotational shear (rad\,d$^{-1}$). The 1$\sigma$ and 2$\sigma$ contours are also shown. Right: The 1$\sigma$ contours from multiple epochs are plotted on top of each other. The mean and uncertainty of the adopted equatorial period and rotational shear are also shown with the crosshair. }
    \label{fig:DiffRot}
\end{figure*}

In Figure \ref{fig:DiffRot} (left panel) we show the reduced-$\chi^2$ landscape for our 2007 May data set, which has excellent phase coverage for tracing the differential rotation of magnetic features (29 observations across 3.87 rotational cycles). The model can be fitted to the noise level, which suggests that the stellar surface is not significantly evolving during the observations. In Figure \ref{fig:DiffRot} the colour represents the reduced-$\chi^2$ fit (with a fixed entropy) for each combination of $\rm{P_{eq}}$ and $\rm{d\Omega}$, and we also show the 1-$\sigma$ and 2-$\sigma$ uncertainty contours around the best-fitting values. In the right panel of Figure \ref{fig:DiffRot}, we show the 1-$\sigma$ uncertainty regions for several of our epochs that have adequate phase coverage to measure differential rotation. The best-fit rotational parameters derived for each epoch vary, which is consistent with the varying parameters derived across previous studies of V889 Her. Given that the uncertainty regions for these epochs overlap, roughly forming a larger uncertainty ellipse, we take these apparent epoch-to-epoch variations as an indication that the uncertainty of our measurement is larger than suggested by the individual 1-$\sigma$ uncertainty regions, similar to \citet{Marsden2007_IMpeg}. For all our models we have adopted the median equatorial rotation period of $\rm{P_{eq}}=$1.328\,d and median rotational shear of d$\Omega=0.319\rm{\,rad\,d}^{-1}$. 

Finally, {\sc{ZDIpy}} uses a maximum degree of harmonic expansion, $l_{max}$, to control the complexity of the magnetic field reconstruction. We tested increasing $l_{max}$, and found there to be no improvement in the reduced-$\chi^2$ fit nor any obvious changes in the reconstructed magnetic field geometry for $l_{max}>18$, therefore we have adopted $l_{max}=18$. 

\begin{table}
    \centering
    \caption{Stellar and line-model parameters for DI and ZDI.}
    \begin{tabular}{lc}
    \toprule
    %Parameter & V889 Her \\ \midrule
   % \multicolumn{2}{c}{Line model parameters} \\ \midrule
   Stellar parameters \\ \midrule
   $\rm{T_{eff}}$ (K) & 5769$^a$ \\
   R ($\rm{R_{\odot}}$) & 1.09$^b$ \\
   Age (Myr) & 30-50$^b$ \\
   Spectral type & G2V$^c$ \\ 
   $\log g$ (dex) & 4.30$^a$  \\ \midrule
   Solar line model parameters \\ \midrule
    $\lambda$ (nm)   &    580$^d$\\
    Gaussian width ($\sqrt{2\sigma}$, km\,s$^{-1}$) &   2.41   \\
    Lorentzian width (half-width at half-maximum, km\,s$^{-1}$) &   1.98    \\
    Effective Land\'e factor    &   1.22$^d$   \\
    Limb darkening coefficient  &   0.66$^e$    \\ \midrule
   % \multicolumn{2}{c}{Fitted stellar parameters} \\  \midrule
   Derived DI/ZDI parameters \\  \midrule
   RV & see Table \ref{tab:results} \\
     $v\sin i$ (km\,s$^{-1}$) & $37.5\pm0.5$\\
     $P_{\rm{rot}}$ (d) & $1.328^{+0.012}_{-0.011}$ \\
     $\mathrm{d\Omega}$ ($\mathrm{rad\,d^{-1}}$) & $0.319\pm0.114$ \\
     Inclination angle ($\degr$) & $60\pm5$ \\
     $l_{max}$ & 18 \\
     \bottomrule
    \end{tabular} \\
    $^a$ \citet{Luck2018}; $^b$ \citet{Strassmeier2003}; $^c$ \citet{Montes2001}; $^d$ \citet{marsden2014}; $^e$ \citet{gray_2005_textbook}
    \label{tab:model_parameters}
\end{table}

\subsubsection{Uncertainty analysis}
We tested the sensitivity of our magnetic field and brightness models to changes in the $\rm{P_{eq}}$, d$\Omega$, RV and $v\sin i$ within their uncertainties. The variation bars shown in Table \ref{tab:results} and Figure \ref{fig:field_properties_timeseries} for the DI and ZDI results relate to the changes in the model outputs while we simultaneously varied the input parameters within their uncertainties. The largest impact comes from variations in the $v\sin i$ which relates to the rotational broadening of the line profile. Changes in the $v \sin i$ are reflected as significant increases in spot coverage. 

\subsection{Chromospheric activity}

We measured chromospheric activity for each RV-corrected observation (Section \ref{sec:RVs}) using the S-index, which is a measure of the excess flux in the cores of the {\ion{Ca}{ii}} H and K absorption lines that is related to non-thermal heating of the chromosphere. S-index observations carried out with NARVAL and HARPS were converted to the common Mount Wilson scale \citep{Wilson1978} using the formula
\begin{equation}\label{eq:Sindex}
    S=\frac{a F_H + b F_K}{c F_{R_{HK}} + d F_{V_{HK}}} +e
\end{equation}

where $F_{H}$ and $F_{K}$ are the fluxes in two triangular bandpasses centred on the cores of the \ion{Ca}{ii} H and K lines (3968.469\,{\AA} and 3933.663\,{\AA})  with widths of 2.18\,{\AA} (at the base), and $F_{R_{HK}}$ and $F_{V_{HK}}$ are the fluxes in two rectangular 20\,{\AA} bandpasses centered on the continuum either side of the H\&K lines at 3901.07\,{\AA} and 4001.07\,{\AA}.  The calibration coefficients a, b, c, d and e for NARVAL data are available in Table 4 from \citet{marsden2014}, and the coefficients for HARPS data were taken from \citet{Gomes2020}.  The wavelength coverage of UCLES spectra do not allow for the S-index to be calculated. 

\subsection{Longitudinal magnetic field}

The longitudinal magnetic field, $B_l$ (G), is the line-of-sight magnetic field strength averaged over the visible stellar surface. For each RV-corrected observation we measured $B_l$ directly from the LSD profiles using the equation
\begin{equation}\label{eq:blong}
    B_{l}=-2.14\times{10^{11}} \frac{\int{vV(v) dv}}{\lambda g c \int{[1-I(v)]dv}}
\end{equation}
where {\it{V(v)}} and {\it{I(v)}} are the Stokes {\it{V}} and {\it{I}} LSD profiles respectively, $\lambda$ (580\,nm) is the central wavelength of the LSD profile, $g$ is the mean Land\'e factor (1.22), c is the speed of light in km\,s$^{-1}$ and $v$ is velocity in km\,s$^{-1}$ \citep{Donati1997}. We integrated over a velocity domain of -52 to 52\,km\,s$^{-1}$ with respect to the line centre to include the entire Stokes {\it{V}} polarization signal while minimizing noise. $B_l$ measurements for each observation are provided as supplementary materials, with error bars determined by propagating the uncertainties computed during the reduction process for each spectral bin of the normalized spectrum through Equation \ref{eq:blong}. 

\subsection{Precise RVs from LSD profiles}\label{sec:RVs}
RVs measured from LSD spectral line profiles can be used to detect exoplanets, as demonstrated by \citet{Lienhard2022}. We measured precise RVs for each individual observation by computing the first-order moment (FOM) of the continuum-normalized Stokes {\it{I}} LSD profile, according to
\begin{equation}\label{eq:FOM}
    \mathrm{RV}=\frac{\int (I_c - I(v))v\, dv}{\int (I_c - I(v)) dv}
\end{equation}
where  $I_c$ is the continuum intensity level, $I(v)$ is the intensity at a velocity $v$ (km\,$\rm{s}^{-1}$), and the integral is computed over a velocity range of -76 to 30\,$\rm{km\,s^{-1}}$ in the heliocentric corrected frame. %Prior to computing the FOM we re-normalized the Stokes {\it{I}} LSD profiles to the continuum, using the inbuilt re-normalization function in {\sc{ZDIpy}}. 
RVs for each individual observation are included in Table 1 of the Supplementary Materials. To determine the uncertainties in each RV measurement, we derived the uncertainty in the FOM by propagation of the uncertainties in the LSD line profile,  and added this in quadrature to the instrumental uncertainty from Section \ref{sec:data_reduction_and_calibration}.% The instrumental uncertainty for UCLES spectra following correction using telluric lines is taken as 0.1$\rm{km s^{-1}}$  \citep{Donati2003_telluriclines}. For NARVAL observations, we adopt an uncertainty of 0.03$\rm{km s^{-1}}$ following telluric correction, based on \citet{Donati2008} and \citet{Moutou2007}.}

\subsection{Photometric rotation period}

The TESS light curves in Figure \ref{fig:TESS} show clear rotational modulation related to brightness contrasts on the stellar surface. The light curve from sector 40 is particularly interesting, with the shape of the modulation changing significantly over $\sim15$ days. The beating and sub-period structure in sector 40 is likely due to rapid evolution of $\geq2$ spots at different latitudes and the presence of significant differential rotation. We measured periods of 1.329\,d through to 1.432\,d for the modulations using the Lomb-Scargle periodogram. These are in reasonable agreement with our adopted equatorial rotation period  and differential rotation rate in Table \ref{tab:model_parameters}, and would reasonably correspond to the differentially rotating equatorial and polar spots. 

\subsection{Photometry smoothing}

The APT differential photometry shows both rotational modulation and significant long-term variability, as has been previously discussed by \citet{Lehtinen2016} and \citet{Willamo2019}. We used a moving average to smooth out short-term variations and observe longer-term periodicities. \citet{Willamo2019} previously took the mean photometric magnitude from a sliding 30\,d window of observations, with a minimum of 14 observations required to be in the window. For the same data presented by \citet{Willamo2019}, we have used a more vigorous smoothing approach to search for brightness fluctuations on a timescale of months to years, taking the mean magnitude from a 180\,d sliding window which is similar in length to an observing season. The raw and smoothed data are shown in Figure \ref{fig:photometry}, and we discuss this further in Section \ref{sec:Longterm_evolution_of_the_magnetic_field_and_brightness_topology}.

\subsection{Activity filtering and an RV search for exoplanets}\label{sec:FilteringActivity}

The stellar RVs we measured from spectral line profiles are a superposition of true centre-of-mass RVs, and apparent spectral line-centre shifts caused by surface temperature contrasts (i.e. dark spots and bright regions) that change the line shape. We modelled these activity-induced RV fluctuations using two methods, the DI method and Gaussian Process Regression (GPR), and searched for underlying periodic RV fluctuations that could indicate the presence of a planetary companion. 

\subsubsection{The `DI method'}

The DI method (e.g. \citealt{Donati_LKCA,Barnes2017}) makes use of the synthetic Stokes {\it{I}} profiles that were generated to model the surface brightness topology of V889 Her (Section \ref{sec:ZDI}), and which give an estimate of apparent RV shifts related to surface brightness features. We measured RVs from both the raw and synthetic LSD profiles using the method described in Section \ref{sec:RVs}, and then subtracted the synthetic RVs directly from the raw RVs. We used the Generalized Lomb-Scargle periodogram \citep{Lomb1976,Scargle1982, Zechmeister2009_LSperiodogram} to search for periodicities in the residual RVs, and assessed the significance of peaks using False Alarm Probabilities (FAPs) calculated using a bootstrap re-sampling approach. Following \citet{Zechmeister2009}, we generated 10000 re-sampled data sets, for each set retaining the timestamps of the RV observations and assigning randomized RV values (with replacement) to each time stamp. We then used the GLS periodogram to find the highest-power peak for each re-sampled data set, and used the maximum peak heights from all 10000 data sets as an estimate of the probability distribution of power maxima. The power values that are above 99.9 and 99 percent of all power maxima are taken as the 0.1 and 1 percent FAPs respectively.

\subsubsection{Gaussian process regression}

The second method was to model the activity-induced RV shifts and their temporal evolution as a Gaussian Process (GP). GPR treats the activity-related RVs as correlated Gaussian noise, where the covariance matrix, {\it{C}}, specifies the correlation between each pair of RV observations. We used the GPR implementation of \citet{Heitzmann2021}, which follows \citet{Haywood2014}, with each entry in the covariance matrix computed from the quasi-periodic kernel
%\begin{equation}
 %   C_{ij}=\theta_1^2 \cdot \rm{exp} \left[ - \frac{(t_i - t_j)^2}{\theta_2^2} - %\frac{\sin^2(\frac{\pi(t_i - t_j)}{\theta_3})}{\theta_4^2} \right] 
%\end{equation}
\begin{equation}\label{eq:GP}
    C_{ij}=\theta_1^2 \cdot \rm{exp} \left[ - \frac{(t_i - t_j)^2}{\theta_2^2} - \frac{\sin^2(\frac{\pi(t_i - t_j)}{\theta_3})}{\theta_4^2} \right] + (\sigma_i^2 + \sigma_s^2)\delta_{ij}
\end{equation}
where the hyper-parameter $\theta_1$ ($\rm{km s^{-1}}$) is the semi-amplitude of the RV signal related to activity, $\theta_2$ (d) is the decay parameter, taken to represent the typical lifetime of stellar surface features, $\theta_3$ (d) is the recurrence time-scale of features ($\sim \rm{P_{rot}}$) and $\theta_4$ is a smoothing  parameter that indicates how complex the signal is within one period. We fixed $\theta_4$ at 0.31 according to \citet{Perger2021}, but we found that having a fixed or variable $\theta_4$ made little difference to our findings in Section \ref{sec:GPR activity modelling}.
%, 0 representing high frequency variations and 1 being low frequency. 
The uncertainty in each data point, $\sigma_i$, and an extra uncorrelated noise parameter, $\sigma_s$, are added in quadrature and applied to the diagonal of the covariance matrix (i.e. the variance of the data). We also include an RV offset (RV$_o$) for each observing instrument. We fitted the GP to the raw RVs using the nested sampling Monte Carlo routine {\sc{pymultinest}} \citep{Pymultinest2014}, a {\sc{python}} implementation of {\sc{multinest}} \citep{Multinest2009}. Priors were taken from Table \ref{tab:GP_priors}. In GPR the rotational spot signal is allowed to change its amplitude and phase over time, which is physically motivated by the short-term changes in mid- to low-latitude spots on V889 Her. Given that the large polar spot on V889 Her is very long-lived, we set the priors for the decay parameter to trace the formation and decay of comparatively short-lived lower latitude features. 

GPR can be used to search for exoplanets by also fitting an activity + planet model to the data, and using Bayesian model comparison to assess the significance of the activity + planet model over the activity-only model. The RV shifts due to an exoplanet in a circular orbit follow the equation
\begin{equation}\label{eq:RVplanet}
    \rm{RV_{planet}}(t)=K \sin\left(\frac{2\pi t}{P_{\rm{orb}}}\right) - \Phi +0.5
\end{equation}
where $\rm{RV_{planet}}(t)$ is the RV shift at a time $t$ (km\,s$^{-1}$), K is the RV semi-amplitude (km\,s$^{-1}$), $P_{orb}$ is the orbital period (d) and $\Phi$ the orbital phase [0, 1]. The Bayesian evidence for each model was computed with the nested sampling MC algorithm during parameter fitting, by integrating over all model parameter spaces according to Equation \ref{eq:bayes_Z}
\begin{equation}\label{eq:bayes_Z}
    Z=\int L(\theta) \pi(\theta) \,d^N\theta 
\end{equation}
where L($\theta$) is the likelihood and $\pi(\theta)$ the prior probability for the set of model parameters $\theta$, and N is the number of model parameters.  According to \citet{Trotta2008} the activity + exoplanet model can be accepted over the activity-only model when the difference in the log of the Bayesian evidence is $\Delta \ln Z > 5$ . 

\begin{table}
    \centering
    \caption{Priors adopted for each of the hyperparameters in equations \ref{eq:GP} and \ref{eq:RVplanet}. $\mathcal{U}$ and $\mathcal{G}$ represent Uniform and Gaussian distributions. Bracketed values are the minimum and maximum for uniform priors, and the mean and standard deviation for Gaussian priors. $\rm{RV_{max}}$ is the maximum unsigned RV, $\rm{\sigma_{RV}}$ is the standard deviation of the RV and T is the time-base of the data set.}
    \begin{tabular}{lr}
    \toprule
    Hyper-parameter & Prior \\ \midrule
    $\theta_1$ (km\,s$^{-1}$)   &   $\mathcal{U}$ (0,$\rm{RV_{max}}$) \\
    $\theta_2$ (d)                  &   $\mathcal{U}$ log(1.33, 100) \\
    $\theta_3$ (d)                  &   $\mathcal{G}$ (1.33, 0.1) \\
    K (km\,s$^{-1}$)              &   $\mathcal{U}$ (0, 2$\rm{RV_{max}}$) \\
    $\rm{P_{orb}}$ (d)              &   $\mathcal{U}$ (0, 0.5T) \\
    $\Phi$                          &   $\mathcal{U}$ (0, 1) \\
    $\rm{RV_{N}}$ (NARVAL)           &   $\mathcal{G}$ (0, $\rm{\sigma_{RV}}$) \\
    $\rm{RV_{H}}$ (HARPS)           &   $\mathcal{G}$ (0, $\rm{\sigma_{RV}}$) \\
    $\rm{RV_{A}}$ (AAT)              &   $\mathcal{G}$ (0, $\rm{\sigma_{RV}}$)\\
    $\rm{\sigma_{N}}$                      &   $\mathcal{G}$ (0, $\rm{\sigma_{RV}}$) \\
    $\rm{\sigma_{H}}$                      &   $\mathcal{G}$ (0, $\rm{\sigma_{RV}}$) \\
    $\rm{\sigma_{A}}$                      &   $\mathcal{G}$ (0, $\rm{\sigma_{RV}}$) \\
    \bottomrule
    \end{tabular}
    \label{tab:GP_priors}
\end{table}
%section{Results and Discussion}\label{sec:Results_Discussion}
\section{Magnetic field and brightness topology of V889 Her}\label{sec:results_discussion1}

Table \ref{tab:results} provides a summary of results from the surface brightness and magnetic field mapping. %, and key results are also plotted in time-series along with the S-index, $B_l$ and RV measurements in Figure \ref{fig:field_properties_timeseries}. The V-band photometry is shown in Figure \ref{fig:photometry}. 
The full set of surface brightness and magnetic field maps are provided in Figures \ref{fig:maps1} through to \ref{fig:maps3}. The fits of the model Stokes {\it{I}} and {\it{V}} line profiles to the observations, and the measured RV, S-index and longitudinal magnetic field strength for each individual observation are also provided as supplementary materials. 

\subsection{Brightness maps}

The bright+spot, and spot-only maps are all dominated by the presence of a large, slightly off-centre and long-lived polar spot (or multiple, unresolved small spots) that extends down to $40-50\degr$ latitude. This is consistent with all previous surface brightness models of V889 Her \citep{Strassmeier2003,Marsden2006,Jeffers2008, Jeffers2011,Willamo2019,Willamo2022}. 
The results also confirm the presence of lower-latitude spots in the visible hemisphere of V889 Her. Previous work by \citet{Marsden2006}, \citet{Jeffers2011} and \citet{Willamo2019} also showed small-scale dark features at the mid- to low-latitudes whereas \citet{Jeffers2008} did not find evidence for such spots in 2005 May to June using data from the MuSiCoS spectrograph \citep{Baudrand1992_Musicos,Donati1999_Musicos} coupled with the TBL, which were unavailable for reanalysis in this work. This difference could be related to the slightly lower resolution of MuSiCoS spectra ($\sim35000$), or it could be that the lower-latitude spots are not always present, although all of our maps show at least weak low-latitude features. It is not clear from our data if there is a relationship between the strength of the polar spot and the existence of lower latitude features, but the mean spot coverage does show some correlation with the strength of the magnetic field (Figure \ref{fig:field_properties_timeseries}), which we discuss further in Section \ref{sec:Longterm_evolution_of_the_magnetic_field_and_brightness_topology}. 

Between the bright+spot and spot-only models, the most striking difference is that the bright+spot maps show weak but significant spot coverage in the `southern' hemisphere of the star. Although there are likely to be spots on both hemispheres, it is not clear if the spots shown in the southern hemisphere are artefacts that compensate for the extensive bright regions in the spot+bright models. The spots in the southern hemisphere are reflected in Table \ref{tab:results} as a significantly larger spot coverage for the bright+spot models compared to the spot-only models. Although we found that the bright+spot models typically converged to lower $\chi^2$ values compared to the spot-only models, this is to be expected for any model with a greater number of variables, and it is not clear which of the inversion techniques results in the most reliable spot model. 

The percentage of the stellar surface covered in spots varies between 3.6 percent in 2008 May and 7.3 percent in 2011 May 14-19, for the spot-only models. Our spot map for 2004 Sept. is similar to that produced by \citet{Marsden2006} using the same data. They obtained a spot coverage of 5.5 percent, which is in reasonable agreement with our 6.9 percent in the spot-only model, considering that \citet{Marsden2006} used a different ZDI implementation from \citet{Brown1991} and \citet{Donati1997b}. 
%\citet{Willamo2019} monitored spot coverage on V889 Her over 18 yrs from 1999 to 2017, but their results cannot be easily compared to ours because \citet{Willamo2019} use comparatively large epochs e.g. 13\,d in their 2010 July data set. Nevertheless, they show fractional spottedness varying between 3.18 and 14.48 percent, which agrees reasonably with our results. 

\subsection{Large-scale magnetic field maps}\label{sec:Magnetic_maps}

The magnetic field maps in Figures \ref{fig:maps1} through to \ref{fig:maps3} indicate a persistent wreath of positive azimuthal field that surrounds the visible pole of V889 Her at all epochs except 2008 May when the magnetic field is at its weakest. Although the weaker field shown in 2008 May may be partly related to the fairly sparse phase coverage of the data, there is similar phase coverage in 2010 Aug., whereas the recovered magnetic field is comparatively strong. Therefore we are confident that the changes in the field strength are not completely related to differences in the phase coverage of data across epochs.  

At latitudes below $\sim60\degr$ the azimuthal field shows mixed polarity regions that have low-to-moderate field strengths (compared to the positive polar wreath). From 2018 June toward the end of our data, the mid-latitudes become dominated by negative regions of azimuthal field, which form a second wreath surrounding the polar wreath. Similar double wreaths persisting over multiple years have been observed for other active stars, like the rapidly rotating K1 subgiant HR 1099 \citep{Petit2004b}, which has a rotation rate approximately 10 times the solar rotation rate (compared to $\sim20$ times the solar rotation rate for V889 Her). Large-scale azimuthal wreaths have also been reproduced in magneto-hydrodynamoic (MHD) simulations of rapidly rotating Suns by \citet{Brown2011}. The presence of such large-scale azimuthal magnetic structures at the stellar surface has been taken as an indication of a fundamentally different dynamo process operating in active stars, compared to that operating within the Sun (e.g. \citealt{Donati2003_telluriclines}).

\begin{figure}
    \centering
    \includegraphics[width=0.9\columnwidth]{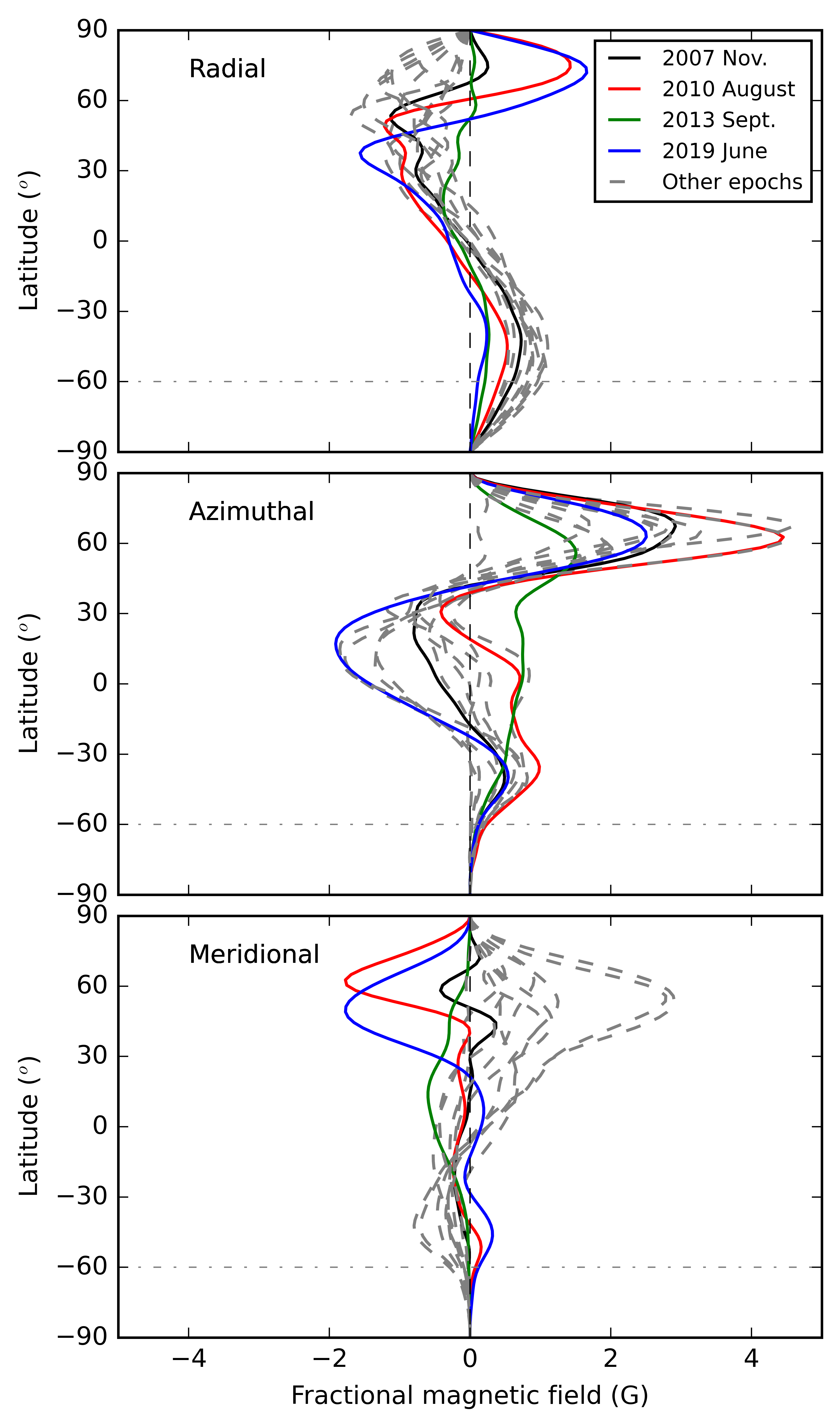}
    \caption{Fractional magnetic field strength per latitude band for the radial (top), azimuthal (middle) and meridional (bottom) field components. For epochs addressed in Section \ref{sec:Magnetic_maps} the data are coloured, and the rest of the epochs are shown as dashed grey lines. The horizontal dashed line at -60$\degr$ latitude indicates the lower extents of our magnetic maps. }
    \label{fig:BvsLat}
\end{figure}

In contrast to the azimuthal field, the radial field shows mixed polarities at high latitudes, consistent with previous observations of V889 Her by \citet{Marsden2006}, \citet{Jeffers2008} and \citet{Jeffers2011}. The radial field over the visible pole and at high latitudes is dominantly negative for most epochs, including in 2011 May 14-15 and 2011 May 17-19, which is consistent with \citet{Willamo2022}. However, the radial field over the pole becomes dominantly positive in 2007 Nov., 2010 Aug., 2013 Sept. and 2019 June, as indicated by Figure \ref{fig:BvsLat}, which shows the fractional magnetic field strength as a function of latitude for each of the radial, azimuthal and meridional field components. The fractional magnetic field was calculated using the formula
\begin{equation}\label{eq:frac_mag_energy}
    F(\theta)=\frac{B(\theta)\cos{(\theta)}d\theta}{2}
\end{equation}
where $F(\theta)$ and $B(\theta)$ are the fractional and average magnetic field strengths of each component at latitude $\theta$, and $d\theta$ is the thickness of individual latitude bands. The switch to a dominantly positive radial field over the visible pole does not indicate a full polarity reversal, with the opposite hemisphere of the star remaining dominantly positive throughout all of our observations. Periods of singular large-scale polarity, where the large-scale magnetic field has the same polarity in both stellar hemispheres, have also been reproduced in the simulations of \citet{Brown2011} for rapidly rotating Suns. 

Interestingly, the dominant field polarity of the meridional component over the pole also changes for 2007 Nov., 2010 Aug. and 2013 Sept. The meridional field polarity even appears to reverse in the hidden hemisphere in 2019 June, but reverts back before 2019 July 10-19. The meridional field is the weakest field component, possibly because ZDI becomes less sensitive to low-latitude meridional fields with increasing stellar inclination \citep{Donati1997b}. %\citet{Donati1997b} found that cross-talk from the meridional field to the radial field can be significant for stellar inclination angles $\geq50\degr$.  This may impact our results most significantly in 2011 May 14-16 and 2011 May 17-19, when the meridional field is at its strongest, and a dominantly negative field is co-located over the visible pole in both the meridional and radial field maps. 

Our magnetic field reconstructions show a stronger magnetic field compared to previously published results from \citet{Marsden2006}, who measured a mean field strength of $\sim30$ G in 2004 Sept. compared to our $\sim87$ G. One reason for this is that we have used different modelling approaches. {\sc{ZDIpy}} is capable of fitting to the Stokes {\it{V}} profiles while taking into account the brightness image, so it can account for regions of reduced flux and increase the magnetic field accordingly. As shown in Figure \ref{fig:Appendix_may01maps}, magnetic models reconstructed in consideration of the brightness topology show considerably stronger magnetic structures, in particular the azimuthal polar wreath, when compared to a magnetic model that does not take into account the brightness topology at all.  Conversely, the field strengths measured here for 2011 May 14-16 and 2011 May 17-19 are an order of magnitude smaller compared to those measured by \citet{Willamo2022}, potentially due to the use of different ZDI software. 

\subsection{Long-term activity and magnetic field evolution}\label{sec:Longterm_evolution_of_the_magnetic_field_and_brightness_topology}

\begin{figure}
    \centering
    \includegraphics[width=\columnwidth]{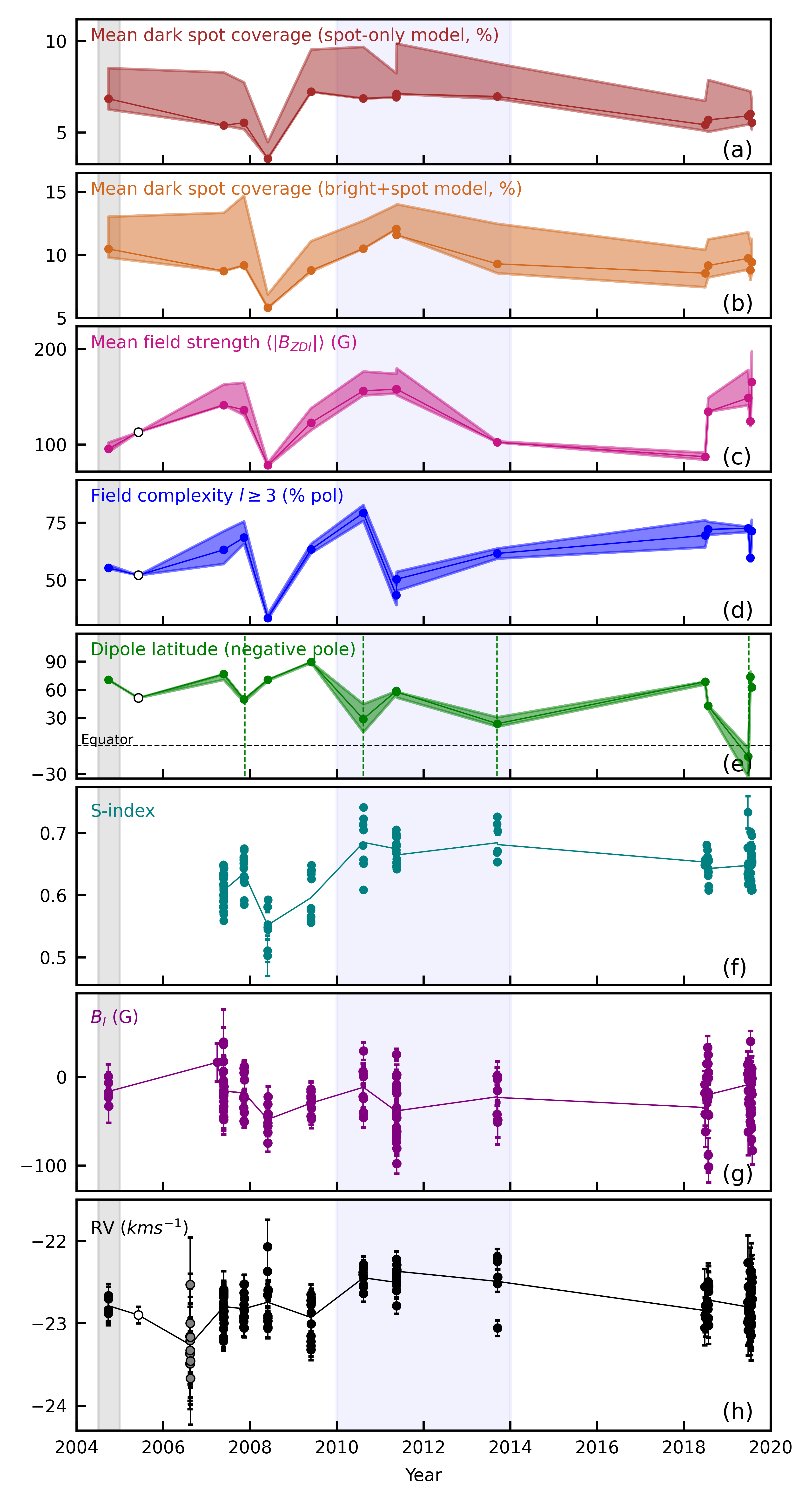}
    \caption{Time-series plots of the surface brightness properties, large-scale magnetic field parameters and magnetic activity metrics for V889 Her between 2004 and 2020. From top to bottom the axes show (a) mean spot coverage for spot-only models, (b) mean spot coverage for bright+spot models, (c) mean magnetic field strength from ZDI, (d) fraction of the poloidal field stored in octopolar and higher-order modes, (e) latitudinal location of the large-scale magnetic dipole (negative pole), with dashed vertical lines indicating possible reversal attempts (f) chromospheric S-index, (g) surface-averaged longitudinal magnetic field strength, and (h) the radial velocity measured from the first-order moment of the observed LSD profile. The shaded regions for plots (a) through (e) indicate the sensitivity of our ZDI results to variations in the $\rm{P_{eq}}$ ($^{+0.012}_{-0.011}\rm{\,d}$), $\rm{d\Omega}$ ($\pm0.114\rm{\,rad\,d}^{-1}$), $v\sin i$ ($\pm0.5$ km\,s$^{-1}$) and RV (see Table \ref{tab:results}). In axes (f) to (h) solid lines join the mean values for each observational epoch. The grey and blue vertical shaded bands indicate data from UCLES and HARPS respectively. Un-shaded data is from NARVAL. Additionally, in (c), (d), (e) and (h) the data shown with  open circles were provided by the authors of \citet{Jeffers2008} (note that these data have no `variation bars'). In (h) the grey filled circles are RVs from \citet{Frasca2010}. }
    \label{fig:field_properties_timeseries}
\end{figure}

\begin{figure}
    \centering
    \includegraphics[width=\columnwidth]{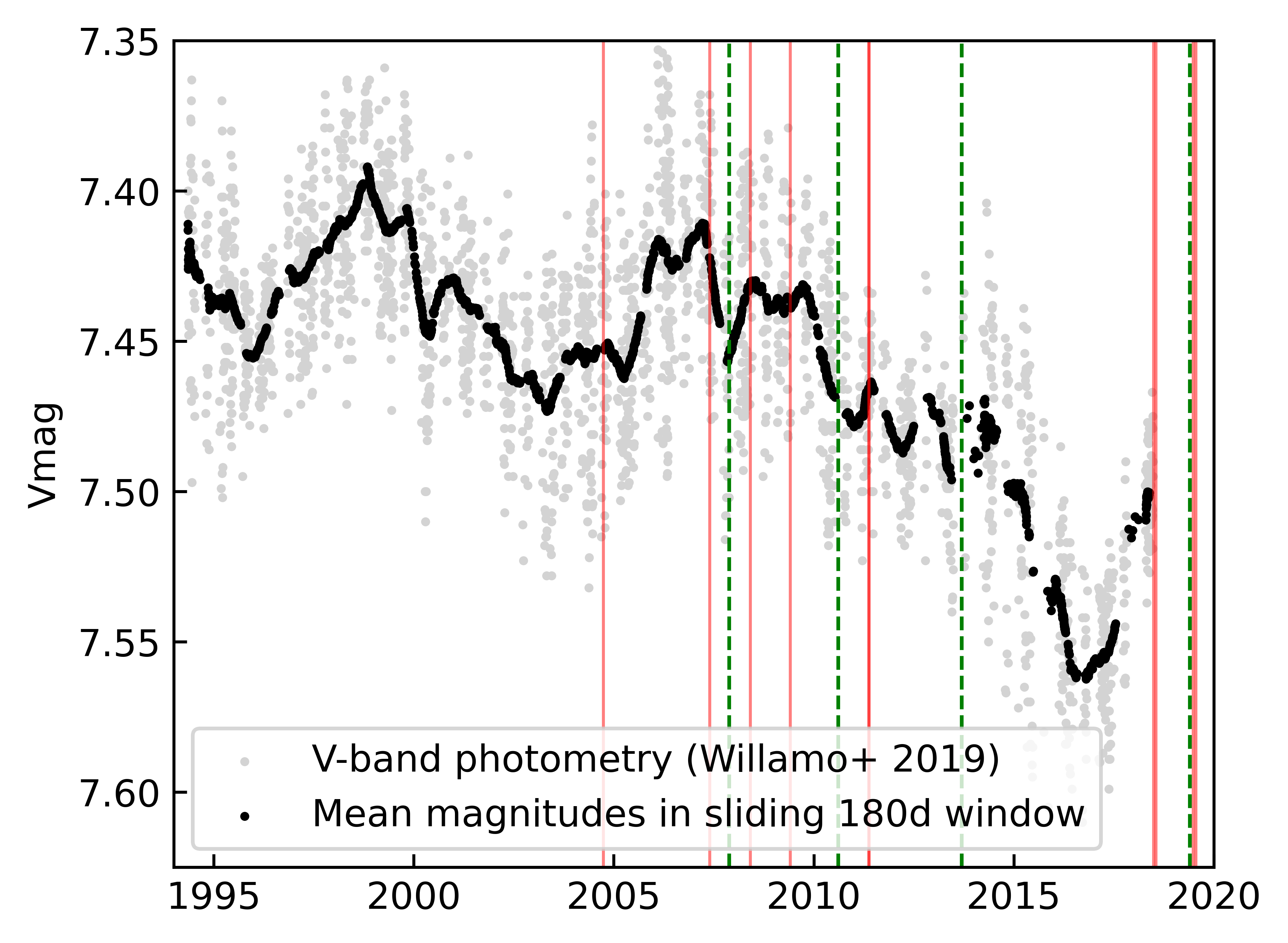}
    \caption{V889 Her brightness variations as shown by differential V-band photometric magnitudes taken between 1994 and 2018. The photometric data, originally published by \citet{Lehtinen2016} and \citet{Willamo2019}, are shown as light grey data points, and our moving 180\,d averages are shown as black points. Green dashed vertical lines indicate the epochs of possible `attempted reversals' in Figure \ref{fig:field_properties_timeseries}, and red vertical lines indicate all other ZDI epochs.}
    \label{fig:photometry}
\end{figure}

Figure \ref{fig:field_properties_timeseries} compares key results from the surface brightness and magnetic field mapping to the chromospheric activity, surface-averaged longitudinal magnetic field strength and radial velocity of V889 Her for the period from 2004 to 2020. The variation bars on the magnetic field parameters in Figure \ref{fig:field_properties_timeseries} were determined by varying the ZDI model input parameters within their uncertainties, and recomputing the brightness and magnetic maps. All of the trends in magnetic field properties that we discuss in this section are more significant compared to the uncertainty bars. We also show the results of our photometry smoothing in Figure \ref{fig:photometry}, where we used a moving 180\,d average to smooth out short-term variations in surface brightness. We have drawn green vertical lines in Figure \ref{fig:photometry} to indicate the epochs where the large-scale magnetic dipole latitude reaches local minima in Figure \ref{fig:field_properties_timeseries}. The synthesis of these data suggests that there are short-term, quasi-periodic magnetic field fluctuations in V889 Her, the first to be observed for such a young solar analogue. 

Between 2007 May and 2011 May 14-16, V889 Her appears to undergo one full oscillation in terms of its magnetic field strength measured from ZDI. The chromospheric S-index and surface-averaged longitudinal magnetic field strength, $B_l$, correlate loosely with the field strength measured from ZDI, with both reaching local minima in 2008 May, and local maxima in 2010 Aug. We determined Pearson's correlation coefficients of 0.52 and 0.39 respectively for S-index and $B_l$ data binned by epoch, with p-values (probability of finding a correlation where there is none) of 0.15 and 0.30.  The magnetic maps also indicate that the field is simple during the magnetic and activity minima in 2008 May, with the fraction of magnetic energy stored in high-order modes ($l\geq3$) at its lowest, while the poloidal field fraction and dipolar field strength are at maxima (Table \ref{tab:results}). As the field strength and chromospheric activity grow toward maxima around 2010/2011, the dipolar component of the field decreases and higher-order components increase as the field structure becomes complex and dominantly toroidal.  Our observations show the negative pole of the large-scale radial dipole dipping toward the equator on two occasions during this period, in 2007 Nov. and 2010 Aug. These epochs are also when the radial field becomes dominantly positive over the visible stellar pole in Figure \ref{fig:BvsLat}. 

The transformation of the magnetic field during this 3-4\,yr period has some similarities to the solar cycle, during which the Sun's magnetic field gains strength and is converted from simple and weak to complex and strong. A key aspect missing for V889 Her, at least in our observations, is the cyclic reversal of the large-scale magnetic dipole, which occurs at around magnetic maxima in the Sun and which regenerates the poloidal field but with opposite polarity. In V889 Her, the large-scale magnetic dipole instead approaches the equator around magnetic maxima, and then the poloidal field is regenerated with the same polarity as before.  MHD simulations by \citet{Brown2011} have showed similar polarity reversal `attempts' for a young Sun rotating at 5 times the solar rotation rate (compared to $\sim$20$\times$ the solar rotation rate for V889 Her). The simulations showed that full reversals can occur in the $5\Omega$-rotating young Sun model, but not every time magnetic energies undergo a full oscillation. There were several failed polarity reversal attempts for each successful one. Further observations will be required to determine whether or not full reversals of the magnetic dipole of V889 Her are possible. 

Although we cannot confirm the presence of any chromospheric activity cycles from our S-index data, at least two potential cycles appear to be present in the smoothed photometry in Figure \ref{fig:photometry}, which we would expect to relate to the formation and decay of spots in a rapidly rotating star such as V889 Her \citep{Radick1990}. There is a long cycle of 7-9\,yrs that is clearly evident in the photometry between 1994 and 2007, and is consistent with the long-term cycle previously reported by \citet{Strassmeier2003}, \citet{Lehtinen2016} and \citet{Willamo2019}. There are also short-term fluctuations in the photometry which occur as rapidly as every $\sim1-1.5$\,yr, around half the period of our observed magnetic field and chromospehric activity fluctuation shown in Figure \ref{fig:field_properties_timeseries}. Interestingly, the magnetic field reversal attempts indicated by our magnetic maps each coincide roughly with local minima in the short-term photometric fluctuations. The fact that the polarity reversal attempts seem to follow photometric cycles supports the presence of repetitive magnetic fluctuations, even though our magnetic field maps have only captured a single fluctuation.

\citet{Jeffers2022} recently showed for $\epsilon$ Eri, which has coexisting $\sim3$\,yr and $\sim13$\,yr chromospheric activity cycles, that the magnetic flux generation and emergence follow the longer chromospheric activity cycle, but are modulated by the shorter cycle. They found that large-scale polarity reversals occur with the longer cycle only, and if V889 Her behaves in the same way this might explain why we have observed no polarity reversals here for V889 Her in phase with its 3-4 yr fluctuations.  The short-term magnetic modulations may be a consequence of a near-surface dynamo. MHD simulations by \citet{Brun2022} with a near-surface dynamo for rapidly rotating stars produced short cycles ($\sim1$\,yr), although these are marked by magnetic polarity reversals, with possible quasi-biennial oscillations. 

It is clear from Figure \ref{fig:field_properties_timeseries} that the magnetic field of V889 Her can also evolve much more rapidly compared to the 3-4\,yr fluctuations, and that the apparent solar-like variability does not strictly hold across all of our observations. In 2019 June, the large-scale dipole of V889 Her crosses over the equator, but still does not fully reverse. The 2019 July 10-19 and 2019 July 20-29 observations seemingly capture the field evolution directly after an attempted reversal, and the magnetic field evolves rapidly during this time. Between 2019 June and 2019 July 10-19, the dipolar axis moves from the equator to 74$\degr$ latitude, and the field strength drops as its structure becomes more simple. $\sim$10 d later, in 2019 July 20-29, the magnetic field has a similar complex structure to 2019 June and has an even stronger field strength, whereas the dipole latitude remains relatively high.  This rapid, non-solar like evolution of the magnetic field could be similar to the chaotic field evolution observed for $\epsilon$ Eri during its short activity cycle. \citet{Metcalfe2013} and \citet{Petit2021} have also noted that $\epsilon$ Eridani potentially alternates between chaotic and cyclic variability. We speculate that the short term evolution of the magnetic field in these young stars could potentially reflect the phase of the observations in the longer activity cycle.  In the work of \citet{Jeffers2022}, $\epsilon$ Eri is approaching a possible grand minimum in activity when its magnetic field is observed to vary chaotically. In our observations of V889 Her, the short-term variability resembles the solar cycle as the star approaches a grand maximum in activity around 2017 \citep{Willamo2019}, inferred from the photometry in Figure \ref{fig:photometry}, and the field evolution is seemingly more chaotic afterwards.

The relationship between spot coverage and the large-scale magnetic field evolution is not entirely clear from our surface brightness and magnetic field maps. Both the spot-only and brightness + spot models suggest that the spot fraction reaches a local minimum in 2008 May, coinciding with minima in the magnetic field strength both from ZDI and the $B_l$, as well as a minimum in the chromospheric S-index.  After 2008 May the spot evolution indicated by the spot-only and brightness + spot models are not consistent, and we cannot be sure which model is the most reliable.  Based on the bright+spot models, a single spot cycle in V889 Her would seem to span two large-scale dipole reversal attempts in 2007 Nov. and 2010 Aug., contrary to the solar case. Given that the photometric brightness varies much more rapidly than this, it is possible that the long spot cycle inferred by our DI results is an effect of infrequent sampling. The presence of both a long-lived polar spot and short-lived low-latitude features in V889 Her also makes deciphering patterns in mean spot coverage and photometric brightness difficult. \citet{Willamo2019} compared spot occupancy to the same (unsmoothed) photometric data we show here. Their spot occupancy data showed rapid variability ($\sim$2-4 yrs between local maxima) %, implying that there would be rapid changes in photometric brightness,  whereas they reported a steadily decreasing trend in brightness from 2007 to 2018. They attributed the lack of agreement to systematic inconsistencies in the determination of the spot occupancy. Conversely, their spot occupancy data 
that would seem to correspond well with the magnetic field and photometric fluctuations shown by our results. For example, their spot filling factors peak between 2009 Sept and 2012 Aug., which is loosely consistent with peaks in the magnetic field strength, complexity and chromospheric activity shown by our observations. The peak amplitude of long-term photometric variations in Figure \ref{fig:photometry} is $\sim$0.09 mag across the epochs of our DI observations, which corresponds to brightness variations of $\sim9$ per cent.  The spot coverage from DI is, as expected, lower than the variability from photometry but is of a similar order.
 
 %%LQ Hya seems similar to V889 Her yet the magnetic behaviour shows stark contrasts. The azimuthal magnetic field does not show prominent rings as we have observed for V889 Her. The radial field of LQHya has been observed to reverse, and this occurred at activity minimum. LQHya is also though to rotate almost as a solid body. LQ Hya \citep{Lehtinen2022} - only weak correlation observed between spotedness and the strength of the azimuthal field component.  They also observed a polarity reversal in the radial field that is linked with a minimum in the spot cycle (max. brightness). Combining maps from \citet{Lehtinen2022} with those from \citet{Donati2003_telluriclines} and \citet{Donati_LQHya1999}, it seems that the polarity remained the same over the entire $\sim20$ yr spot cycle up until the polarity change at spot minimum. Interestingly, there is no differential rotation and the azimuthal wreaths we have reconstructed do not appear to be present for this star \citet{Donati2003_telluriclines,Lehtinen2022}.

\subsection{Differential rotation}
We measured a high level of differential rotation from several of our magnetic field maps, as shown in Figure \ref{fig:DiffRot}. For epochs with results not shown in Figure \ref{fig:DiffRot}, which have poorer phase coverage and sampling frequency, the $\chi^2$ minimization tests either didn't converge to a clear minimum, or we measured zero rotational shear. We consider these to be less reliable measurements. Modelling by \citet{Brown2011} suggested that differential rotation would change throughout a magnetic cycle due to dynamo waves, and observations of some cool stars support this \citep{Petit2004}, but it is not clear from our data if the evolution of the differential rotation parameters for V889 Her follow any trend with respect to the magnetic fluctuations. Like \citet{Marsden2007_IMpeg} did for IM Peg, we take the apparent epoch-to-epoch variations in the rotational shear of V889 Her, and the fact that the 1-$\sigma$ uncertainty regions from different epochs overlap, as an indication that the uncertainty of our measurement is larger than suggested by individual epochs.

\section{The search for exoplanets orbiting V889 Her}\label{sec:results_discussion2}

\subsection{RV variations on both short and long time-scales}\label{sec:RVs_discussion}

The RV as measured from the Stokes {\it{I}} LSD observations shows both short-term fluctuations that relate to the rotation of magnetic surface features in and out of view, and also longer-term RV variability, the nature of which is unclear. 

\begin{figure}
    \centering
    \includegraphics[width=\columnwidth]{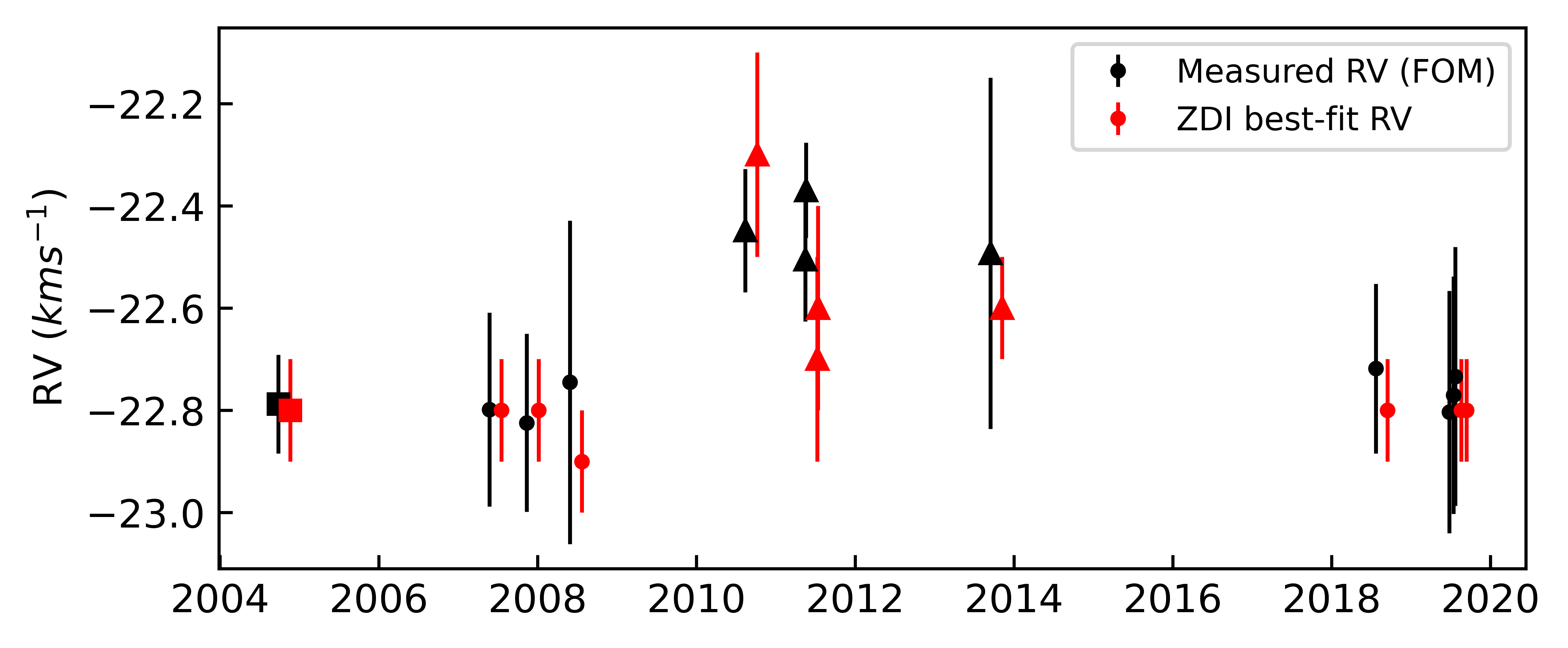}
    \caption{Comparison of the observed mean RV (black) for each DI epoch to the line-centre RV adopted to optimize the DI model fit (red, taken from Table \ref{tab:results}). Squares, circles and triangles indicate data from UCLES, NARVAL and HARPS respectively.  The bars for the measured RVs show the standard deviation of the individual observations, and for the modelled line-centre RVs the bars represent the uncertainties in the best-fit line-centre RVs. The data in red have been shifted to the right for clarity.}
    \label{fig:RV_comparison}
\end{figure}

\begin{figure}
    \centering
    \includegraphics[width=\columnwidth]{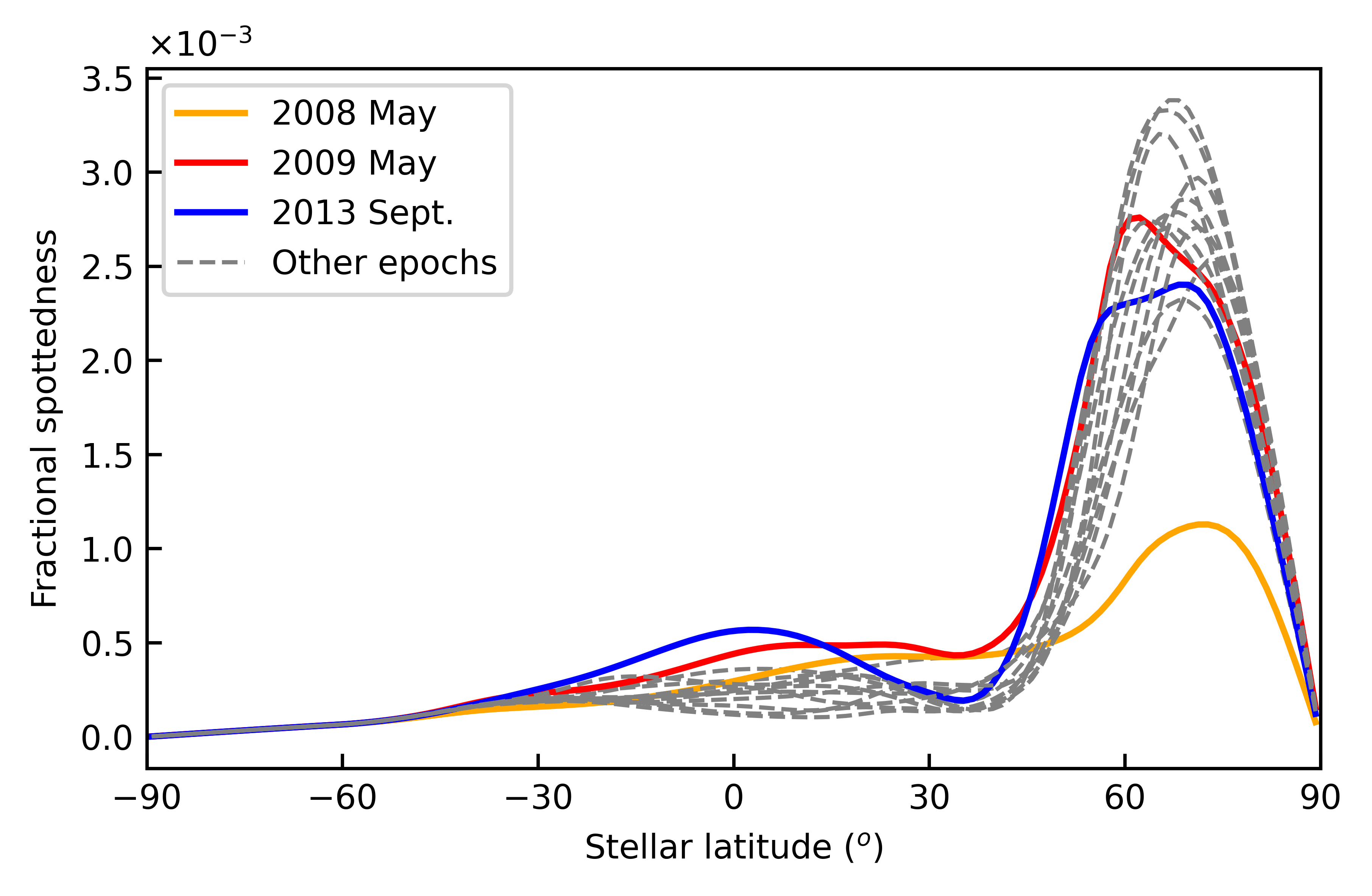}
    \caption{Fractional spot coverage versus stellar latitude for all of our DI maps. For epochs addressed in Section \ref{sec:RVs_discussion} the data are coloured, and the rest of the epochs are shown as dashed grey lines. }
    \label{fig:frac_spot}
\end{figure}

In Figure \ref{fig:RV_comparison} we show the mean and standard deviation of the observed RVs from Figure \ref{fig:field_properties_timeseries} for each of our DI epochs (black markers). The standard deviation of the short-term RV variations is the greatest in 2013 Sept. This is also when the fractional spottedness at the equator is the greatest, according to Figure \ref{fig:frac_spot} which shows the variation in fractional spottedness with stellar latitude for each of our DI models. The fractional spottedness was calculated in the same way as the fractional magnetic field strength, using Equation \ref{eq:frac_mag_energy}, but with the average field strength at a latitude $\theta$ ($B(\theta)$) replaced by the average spot occupancy. Figure \ref{fig:frac_spot} also shows high spot fractions at the mid-to-low latitudes in 2008 May and 2009 May, and the RVs in Figure \ref{fig:RV_comparison} show large standard deviations for these epochs. These data suggest that the amplitude of rotational RV modulations in V889 Her is most impacted by low-latitude surface features.

The short term RV oscillations are not about a consistent RV, but rather a slightly varying RV across our observational epochs. This is consistent with the long-term RV variations observed over several months to years by \citet{Huber2009}. For comparison against the observed RV in Figure \ref{fig:RV_comparison}, we also show in red the line-centre RV that optimized the DI model fit for each of the epochs.  The apparent line-centre RV variations in Figure \ref{fig:RV_comparison} are not sufficiently larger than their uncertainties for us to believe they are true RV fluctuations, nor are the data adequate to carry out a period search, but we have adopted the fluctuating RVs to optimize the DI and ZDI model fits. In 2011 May 17-19 the offset between the observed and the best-fit line-centre RV is the most significant, and this is also when the spot occupancy is the greatest in Figure \ref{fig:field_properties_timeseries}. Apart from this, there is no clear relationship between the two data sets in Figure \ref{fig:RV_comparison}, and it is not clear if the long-term RV shifts can be explained by the magnetic activity evolution of V889 Her, are indicative of a long-period companion, or a combination of the two. The HARPS data all show mean measured RVs that are higher compared to the AAT and NARVAL data, but we do not expect that the long-term RV fluctuations can be explained by an instrumental offset, because we corrected for the heliocentric velocity of the observatory toward the star and used telluric lines to reduce night-to-night shifts in the NARVAL and UCLES RVs, as discussed in Section \ref{sec:data_reduction_and_calibration}. 

\subsection{DI activity modelling}\label{sec:DI_activity_modelling}

\begin{figure*}
    \centering
    \includegraphics[width=0.50\linewidth]{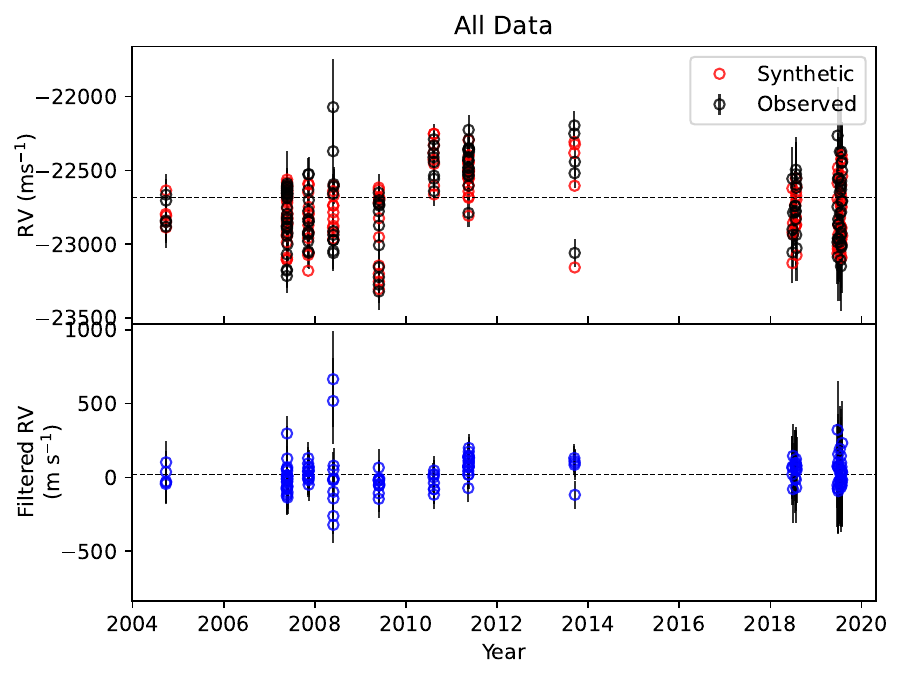}
        \includegraphics[width=0.495\linewidth]{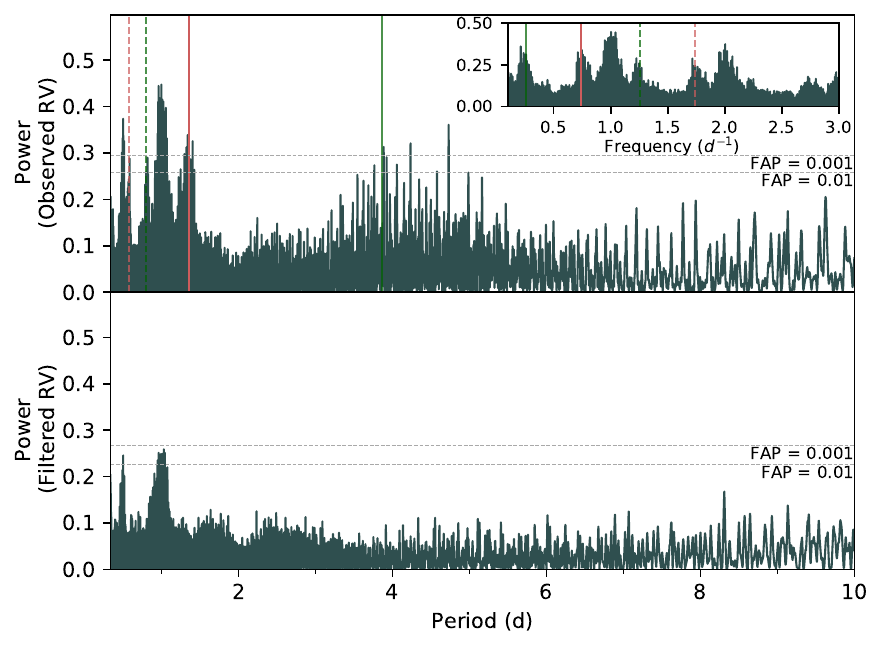}
    \caption{Left: Observed (black) and synthetic (red) RVs of V889 Her for combined data from all DI epochs (top) and residual RVs calculated by subtracting the synthetic from the observed RVs (bottom). %The horizontal axis gives the date with respect to the HJD 2458690.0, which corresponds to a rotational cycle of 0 in the 2019 July 20-29 data set.     Uncertainties in the observed RVs were derived by propagating the uncertainties in the Stokes {\it{I}} LSD profile through Equation \ref{eq:FOM}.
    Right: Power spectra for the observed (top) and residual (bottom) RVs, computed using the GLS periodogram. The solid red vertical line indicates the stellar rotational signal, and the solid green line indicates its alias at a frequency of $1-\frac{1}{\rm{P_{rot}}}$. Aliases of both of these signals repeat at frequency intervals of 1\,$\rm{d^{-1}}$, and are indicated by the dashed red and green lines. Inset, we also include the power spectrum of the observed RVs in the frequency domain, where the aliasing of the rotational signal is more clear. The horizontal dashed grey lines indicate the periodogram power levels that correspond to a 0.1 percent and 1 percent FAP.}
    \label{fig:DI_RV_filtering}
\end{figure*}

Figure \ref{fig:DI_RV_filtering} (top panel) shows the observed RVs and our synthetic, activity-induced RVs from DI, for a combination of data from all DI epochs. The `filtered' RVs are shown in the bottom panel of Figure \ref{fig:DI_RV_filtering}, which are the residual values after subtracting the synthetic RVs from the observed RVs. DI filtering results in a 136 m\,s$^{-1}$ improvement in the RV RMS (raw RVs have RMS $\sim250$ m\,s$^{-1}$ compared to $\sim$114\,m\,s$^{-1}$ for filtered RVs). Power spectra for the observed and residual RVs are also shown in Figure \ref{fig:DI_RV_filtering}, where significant frequencies in the data may correspond to a higher periodogram power.  The power spectrum for the observed RVs is dominated by forests of peaks around the 1\,d observing cadence and its aliases (0.5\,d, 0.25\,d etc.). There are also forests of strong peaks around the rotation period (marked with a red line), an alias of the rotational signal at a frequency of $1-\frac{1}{\rm{P_{rot}}}$ d$^{-1}$ (marked with a green line) which is related to the observing cadence, as well as aliases of both of these signals that recur at frequency intervals of $\sim$1 d$^{-1}$ (dashed red and green lines). The power spectrum for the filtered RVs shows residual peaks at periods of 1\,d and 0.5\,d, related to the observing cadence. The results are similar for all other data sets we tested. The results for other selected data sets are provided as supplementary materials. 

Although the power spectra in Figure \ref{fig:DI_RV_filtering} prove we were able to remove the activity-related rotational RV signal and its aliases by subtracting the synthetic RVs, we were still unable to recover any significant periodicities in the residuals outside those related to the observing cadence, and this is an impact of the limited temporal coverage of our data.  %For some epochs the residuals appear to show potential trends, such as 2011 May 14-16 and 2011 May 17-19 (see Figure \ref{fig:DI_filtering_results1}) when the residual RV climbs smoothly over $\sim6$\,d, but our data are not adequate to determine if these could be periodic variations. 
For V889 Her, longer periods of dense observations could improve exoplanet detection capabilities, but care will need to be taken when using the DI activity filtering technique because of the underlying long-term RV variability. We are not sure of the nature of the long-term RV shifts, but the varying line-centre RVs adopted for each DI epoch undoubtedly bias the DI activity filtering and thus the exoplanet search. The impact would likely be the greatest when searching for long-period companions, such as planets orbiting within the habitable zone. 

\subsection{GPR activity modelling}\label{sec:GPR activity modelling}

We also tested GPR to model the magnetic activity of V889\,Her, using a quasi-periodic GP kernel which has been shown to be highly effective at reproducing rotationally modulated activity-induced RVs (e.g. \citealt{Haywood2014,Angus2018,Heitzmann2021,Perger2021}). However, our GPR models did not converge to a reasonable solution for any of our activity-only or activity + exoplanet tests, so here we explore the challenges of applying GPR to V889 Her as a test case for young solar analogues. 

\subsubsection{Issue 1: Different decay time-scales for polar and lower-latitude spots}

A key issue was constraining the hyper-parameter $\theta_2$ which relates to the time-scale for active region growth and decay. When we applied GPR to single epochs of data $\theta_2$ did not converge to a realistic solution (Figure \ref{fig:2007may_GPresults}). Potentially the RV signal related to the long-lived polar spot obfuscates the signal from lower-latitude features, which we would expect to decay rapidly over $\sim3$ to $15$\,d based on our brightness and magnetic maps, and the TESS light curve in Figure \ref{fig:TESS}. The polar-spot evolves on a timescale much longer than a single epoch of data and \citet{Nicholson2022} found that $\theta_2$ breaks down when this occurs. %To better constrain the decay timescale for lower-latitude spots, future observations should be carried out for as long as possible at high cadence, covering at least 15\,d. 

When combining data from multiple epochs the GP could fit to either short (several days) or long (several thousand days) spot decay timescales depending on the prior used for $\theta_2$ (see Figure \ref{fig:HARPS_GPresults}). There was not a significant difference in the log of the Bayesian evidence between these models, so the preferred model is unclear. 

\subsubsection{Issue 2: Strong differential rotation}

When fitting to multiple epochs of data the posterior distribution of the rotation period, $\theta_3$, seemed to be correlated with the decay timescale. A slightly longer rotation period was associated with a longer spot decay timescale (likely related to the polar spot), and vice versa.  When we fitted the GP to the entire $\sim15$\,yr of data, $\theta_3$ even showed a bimodal distribution (Figure \ref{fig:AllData_GPresults}). This is probably related to the significant differential rotation exhibited by V889\,Her, which means that the RV modulations caused by spots at different latitudes will have different periods. %, with periods of $\sim1.34$\,d and $1.37$\,d being equally likely.  

\subsubsection{Issue 3: Insufficient temporal coverage of data}

Given the above 2 issues, the temporal coverage of our legacy data is insufficient to model the activity-induced RVs for V889 Her using GPR. Our data is not overall very different from that used by \citet{Heitzmann2021} (23 UCLES spectra covering 11\,d or 4.68 stellar rotations) but they found that a quasi-periodic GP could satisfactorily reproduce the activity-induced RVs of HD\,141943 for a single observational epoch.  Their target has a rotation period roughly double that of V889 Her, at 2.198$\pm$0.002\,d, and lower differential rotation rate at $d\Omega=0.1331^{+0.0095}_{-0.0094}$\,rad\,d$^{-1}$ compared to 0.319\,rad\,d$^{-1}$ for V889 Her (although they measured DR from the Stokes {\it{I}} profiles while we used Stokes {\it{V}} profiles). Their reconstructed DI images indicate a polar spot that appears to be somewhat smaller compared to the polar spot on V889 Her, and they recovered low-latitude features with a similar strength to the polar spot and much larger area coverage compared to in V889 Her, for which the low-latitude features are weak and small relative to the polar spot.  The strength of the lower latitude features relative to the polar spot, and the low differential rotation rate in HD\,141943, may explain why the GP converges easily for the moderately active star in \citet{Heitzmann2021}. 

Another issue in modelling magnetic activity for V889\,Her is the underlying long-term RV variations and the fact that our observations with different instruments have no temporal overlap. The mean RV from HARPS is $\sim300$\,m\,s$^{-1}$ higher compared to the AAT and NARVAL RVs, differences that the GP can easily account for with large instrumental offsets, but this dilutes the contributions of activity and any exoplanets. Given that we corrected for the heliocentric velocity of the observatory toward the star and used telluric lines to reduce night-to-night shifts in the NARVAL and UCLES RVs, we do not expect such large RV differences between the data sets to be entirely instrumental.  

It is clear from our results that DI is more suitable compared to GPR for modelling the magnetic activity of V889\,Her using our current data set. DI has more flexibility to model complex magnetic activity, compared to our quasi-periodic GP model. However, DI cannot resolve small scale magnetic structures, which \citet{Jeffers2014b} and \citet{Lisogorskyi2020} showed can prevent the detection of up to 20 $M_{\oplus}$ planets in temperate zones around Sun-like stars. Therefore it is important that we improve data-driven techniques such as GPR to model the magnetic activity of young Suns. 

\subsubsection{Recommendations for modelling extreme magnetic activity with GPR}

Although \citet{Perger2021} found that a quasi-periodic GP can fit well individually to solar-like low-latitude spot configurations, or polar spots, our results suggest that the quasi-periodic GP may not be able to constrain the activity evolution for a combination of the two when significant differential rotation is present, at least when using legacy data with large gaps. Future observing plans should focus on obtaining long time-series of continuous high-cadence observations, ideally with nightly observations spanning a few months as is carried out with the RedDots exoplanet search program \citep{Jeffers2020_RedDots}.  Exploration of alternative GP covariance kernels and simultaneous modelling of RVs and activity indicators are also recommended. We performed some preliminary tests combining a quasi-periodic and stable periodic GP component \citep{Kossakowski2022} to model contributions of the low-latitude and polar features respectively to the activity-induced RVs, but the GP did not converge for our tests. 

\subsection{Exoplanet detection limits with DI}

%Following \citet{Zechmeister2009} and \citet{Bonfils2013}, we derived period-mass limits above which we can exclude the presence of a planet orbiting V889 Her, given our observations. We used the DI activity-subtracted RVs from Section \ref{sec:DI_activity_modelling}. As in \citet{Zechmeister2009}, we used bootstrap re-sampling on the activity-subtracted RVs to estimate the periodogram power-FAP relation for a no-planet scenario. 

We used V889 Her to estimate the detection limits for HJs orbiting extremely active stars when using existing legacy data and filtering activity jitter using DI. We injected synthetic planetary RV shifts into our data, re-computed our DI models and then searched the DI-subtracted RVs for the injected planetary signals. The planet-induced RV shifts were calculated using equation \ref{eq:RVplanet}, assuming a circular orbit, and the shifts were applied to the observed LSD line profiles. We limited our tests to planets with orbital periods of 3 to 7\,d (orbiting at $\sim$0.04 to 0.07 AU), and masses of 1 to 2 $M_{Jup}$. 

\begin{figure}
    \centering
    \includegraphics[width=\columnwidth]{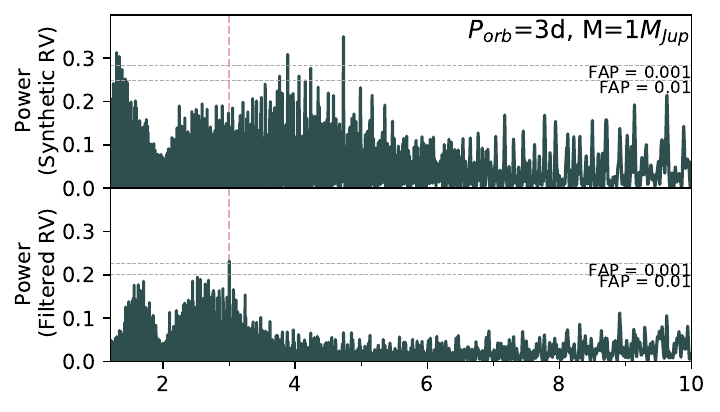}
    \includegraphics[width=\columnwidth]{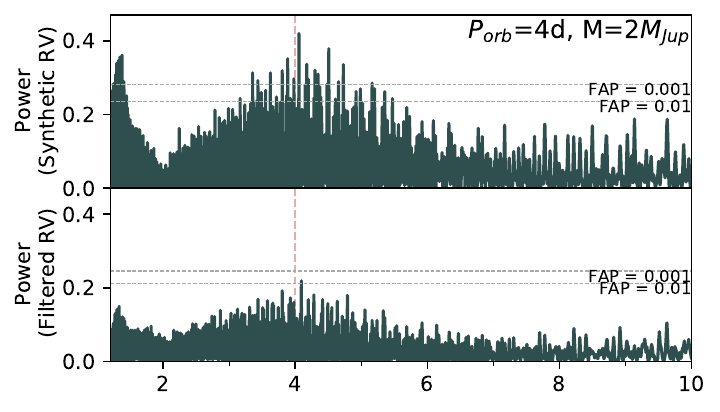}
    \includegraphics[width=\columnwidth]{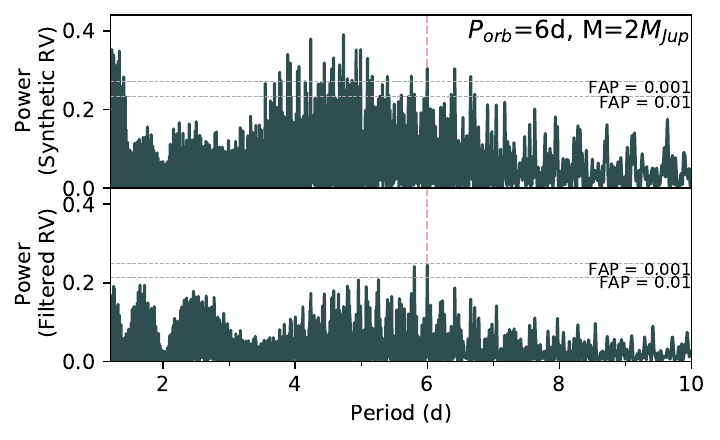}
    \caption{Power spectra for synthetic and DI-subtracted RVs (all epochs combined), where the RVs include a synthetic planetary signal with a period and planetary mass as indicated in the Figure. The dashed red vertical line in each panel indicates the period of the injected planet. The horizontal dashed grey lines indicate the periodogram power levels that correspond to a 0.1 and 1 percent FAP, which were computed as in Section \ref{sec:DI_activity_modelling}. We have excluded periods $\leq1.2$\,d, for which the power spectrum is dominated by peaks related to the observing cadence and  its aliases.}
    %Minimum detectable mass versus orbital period for exoplanets in a circular orbit around V889 Her. The detection limit shown by the black line is the 99.9 percent confidence limit for detection in the DI activity-subtracted RVs for 2007 May (top), combined data from 2011 May 14-16 and 2011 May 17-19 (middle), and all data (bottom). We include horizontal grid lines at different masses for clarity. }
    \label{fig:Detection_limits}
\end{figure}

Given our $\sim15$\,yr data set we estimate that we could, in principle, detect Jupiter-mass planets with orbital periods of 3\,d.  Figure \ref{fig:Detection_limits} (top) shows the power spectrum for the entire series of RVs with the injected 3\,d, 1$M_{Jup}$ planetary signal, and the power spectrum of the DI-subtracted RVs. The DI-subtracted RVs show a 3\,d peak which is significant with respect to the 0.1 percent FAP level.  %We also carried out the period search using shorter series' of data, such as the subset of HARPS observations, and although the power spectrum showed a forest of peaks around 3\,d, the peaks were not significant with respect to the 0.001 FAP level. This shows the benefit of a longer timeserie}

At orbital periods of 4-5\,d the planetary signal is buried under a forest of peaks with periods around $1/(1+f_{rot})$, which are aliases of the stellar rotational signal. The rotational signal and its aliases are removed by DI such that 2\,$M_{Jup}$ planets at these orbital periods would be detectable with a FAP $<1$ percent, as shown in the second panel of Figure \ref{fig:Detection_limits}. 2\,$M_{Jup}$ planets with orbital periods of 6-7\,d are detectable in the DI-subtracted RVs with a FAP $<1$ percent. However, for these longer orbital periods there are several strong signals shown in the DI-subtracted RVs and selecting the correct peak is non-trivial. This issue arises as the orbital period becomes similar to the length of most our individual epochs of observations.

%\begin{figure}
 %   \centering
 %   \includegraphics[width=0.8\columnwidth]{Figures/Detection_limits_conservative.png}
  %  \caption{Caption}
  %  \label{fig:Detection_limits}
%\end{figure}

\section{Conclusions}\label{sec:conclusions}

In this study we used previously published and new spectropolarimetric data spanning the years 2004 to 2019, with archival photometry taken between 1994 and 2018, to investigate the variability of the magnetic field, surface activity and RV of the young solar analogue V889 Her. We also tested the techniques of Doppler Imaging (DI) and Gaussian Process regression (GPR) to filter activity-induced RV variations and search the residual RVs for exoplanets. 

Our magnetic maps indicate possible quasi-periodic magnetic field fluctuations with a potential period of 3-4 yr. The magnetic field evolution during this period loosely resembles the solar magnetic cycle; the field is simple and dominantly poloidal during magnetic and chromospheric activity minima, and is complex and dominantly toroidal during a magnetic and chromospheric activity maxima.  Smoothed photometric data show a 7-9 yr brightness cycle and short-term brightness oscillations with a period of 1-1.5 yr, roughly half the length of the magnetic field oscillation we observed.  
Contrary to the solar case, no large-scale polarity reversals were observed for V889 Her, but several attempts at polarity reversals were observed. Long-term monitoring will be required to determine if the star undergoes large-scale magnetic polarity reversals and if it truly switches between solar-like and non-solar like behaviour.  
 
Our surface brightness maps confirmed the presence of both a large, long-lived polar spot and multiple smaller short-lived spots at low- to mid-latitudes. The observed stellar RV showed significant rotational modulation, its amplitude tied to the spot coverage at low latitudes. The RVs also showed underlying long-term fluctuations of unconfirmed origin. 

We were unable to reliably fit a quasi-periodic GP model to the stellar activity-induced RV variations due to three main challenges:
\begin{enumerate}
    \item There are competing RV signals from the long-lived and temporally varying polar spot and rapidly evolving low-latitude features; 
    \item There is significant differential rotation; and
    \item The temporal coverage of our legacy data is insufficient.
\end{enumerate}
To help overcome these challenges we recommend that alternative GP kernels be tested and high cadence observations of V889 Her (or similar stars) be carried out for as long as possible, ideally with nightly observations spanning a few months as is carried out with the RedDots exoplanet search program \citep{Jeffers2020_RedDots}.

When using existing legacy data, the DI technique can be a useful tool for modelling the activity-induced RV variability of active young Suns, with the flexibility to model complex surface spot distributions. Using the DI method we were able to remove the activity-related rotational RV signal and its aliases. Given our $\sim15$\,yr set of RV observations we estimate that we could, in principle, detect the presence of a 1\,${M_{Jup}}$ planet with an orbital period of $\sim3$\,d and, with less certainty, 2\,${M_{Jup}}$ planets with orbital periods between 3 and 7\,d.

\section*{Acknowledgements}
We thank the referees for their constructive feedback, which has helped us to improve the quality of this manuscript.

This work is based on spectropolarimetric observations obtained at the TBL, AAT and 3.6-m ESO telescope. We thank the technical staff at each of these facilities for their time and data. We also acknowledge the use of the PolarBase database, which makes TBL observations publicly available, and is operated by the Centre National de la Recherche Scientifique of France (CNRS), Observatoire Midi-Pyrénées and Université Toulouse III - Paul Sabatier. 
%The TBL is operated by the Institut National des Sciences de l'Univers of the .  We thank the staff at the TBL for their time and data. 
We acknowledge use of the {\sc{simbad}} and {\sc{VizieR}} data bases operated at CDS, Strasbourg, France. This work has also made use of the VALD database, operated at Uppsala University, the Institute of Astronomy RAS in Moscow, and the University of Vienna.
ELB is supported by an Australian Postgraduate Award Scholarship. SVJ acknowledges the support of the German Science Foundation (DFG) priority program SPP 1992 `Exploring the Diversity of Extrasolar Planets' (JE 701/5-1). 
In addition to the {\sc{python}} packages referenced in Section \ref{sec:Analysis}, this research has made use of the following packages:
{\sc{astropy}} \citep{Astropy2013}, {\sc{corner}} \citep{Corner2016}, {\sc{loguniform}} (MIT licence; Joao Faria), {\sc{matplotlib}} \citep{Matplotlib2007}, {\sc{NumPy}} \citep{NumPy},  and {\sc{SciPy}} \citep{SciPy}.

%%%%%%%%%%%%%%%%%%%%%%%%%%%%%%%%%%%%%%%%%%%%%%%%%%
\section*{Data Availability}

Spectropolarimetric data are available via PolarBase (http://polarbase.irap.omp.eu/), the ESO archive (http://archive.eso.org/cms.html) and AAT archive (https://archives.datacentral.org.au/). TESS data can be found at the Barbara A. Mikulski Archive for Space Telescopes (MAST, https://mast.stsci.edu/), and the differential V-band photometry is available on {\sc{VizieR}} \citep{Vizier:Willamo2019}. 

%%%%%%%%%%%%%%%%%%%% REFERENCES %%%%%%%%%%%%%%%%%%

% The best way to enter references is to use BibTeX:

\bibliographystyle{mnras}
\bibliography{library} % if your bibtex file is called example.bib

\appendix

\section{TESS photometry}

\begin{figure*}
    %\centering
    \includegraphics[width=0.8\linewidth]{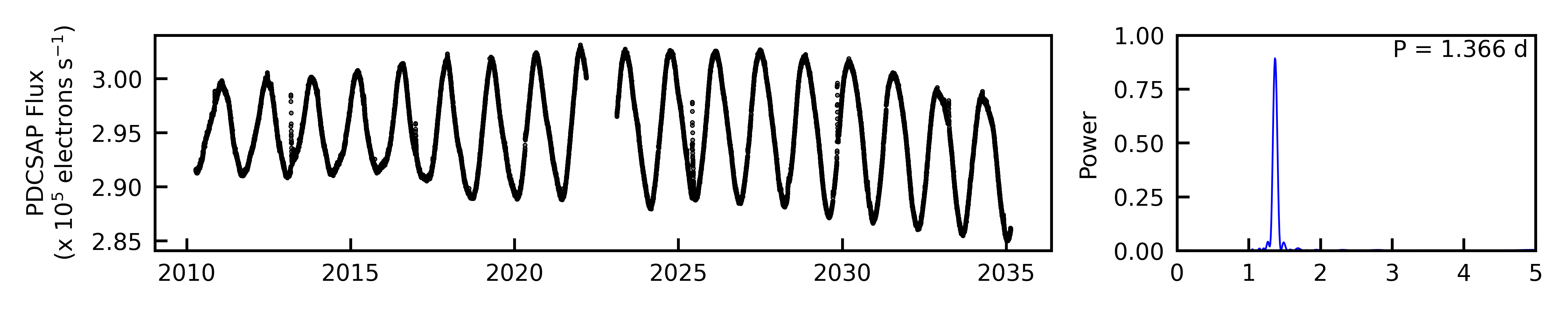}   
    \includegraphics[width=0.8\linewidth]{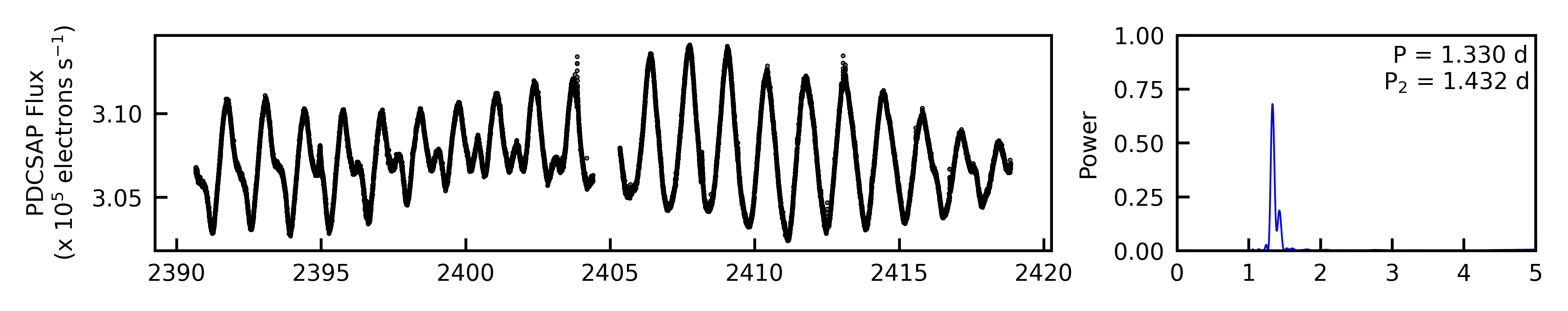}
    \includegraphics[width=0.8\linewidth]{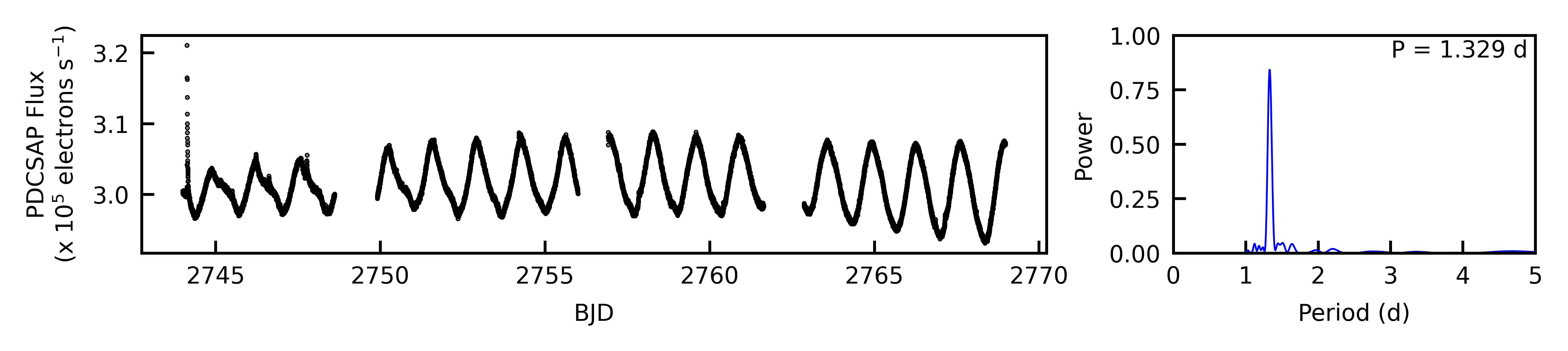}
    \caption{TESS Pre-search Data Conditioned Single Aperture Photometry (PDCSAP) observations of V889 Her from sectors 26 (top), 40 (middle) and 53 (bottom). The data show clear rotational modulation related to brightness contrasts on the stellar surface. Occasional vertical lines in the flux are indicative of flares. Power spectra derived using the GLS periodogram are also shown on the right of each data set, and the period of the brightness modulation is indicated as P. For sector 40 (middle) P$_2$ is the period related to the secondary power spectrum peak. }
    \label{fig:TESS}
\end{figure*}

Figure \ref{fig:TESS} shows TESS photometry from sectors 26, 40 and 53, where V889 Her was observed in 2-minute integrations. Power spectra derived using the GLS periodogram \citep{Zechmeister2009_LSperiodogram} are also shown for each data set.  

%\begin{figure}
 %   \centering
 %   \includegraphics{Figures/TESS_timeseries.pdf}
  %  \caption{TESS PDCSAP observations of V889 Her from sector 26 (red) with a 1.366\,d sinusoid (black) fitted using the GLS periodogram \citep{Zechmeister2009_LSperiodogram}.}
  %  \label{fig:TESS_rotation_period}
%\end{figure}

\section{ZDI results}

\begin{figure*}
    \centering
    \includegraphics[width=\linewidth]{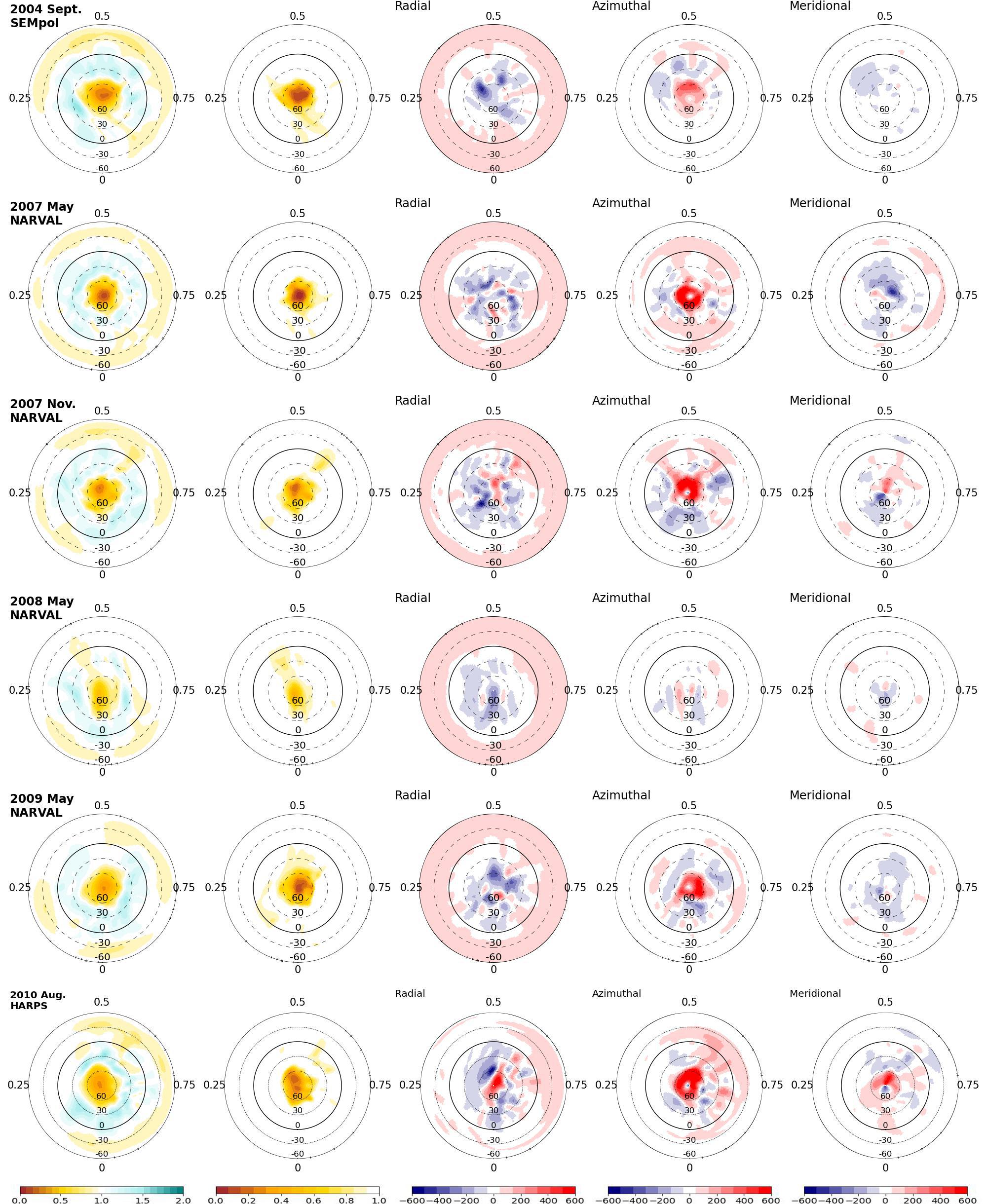}
    \caption{Left to right: reconstructed brightness + spot, spot-only, radial field, azimuthal field and meridional field maps for V889 Her, from 2004 Sept. through to 2010 Aug. The maps are flattened polar projections, showing the visible pole at the centre and with latitudes extending down to -60$\degr$. Radial ticks indicate the phases of observations. }
    \label{fig:maps1}
\end{figure*}

Table \ref{tab:results} summarizes the ZDI model outputs, and Figures \ref{fig:maps1} through to \ref{fig:maps3} show the brightness and magnetic field maps for each of our 14 observational epochs. Model fits corresponding to each set of maps are shown in Figures 1 to 4 of the supplementary materials.

\setlength{\tabcolsep}{3pt}
\begin{table*}
\caption{Surface brightness and large-scale magnetic field properties for 2004 Sept. through to 2019 July. Columns (left to right) indicate the best-fit RV for each epoch, the mean percentage of the stellar surface covered by dark spots for a spot-only model ([0:1]), the mean spot coverage ([0:1]) and mean surface brightness (relative to a quiet photosphere with 100 percent brightness [0:2]) for a bright+spot model, unsigned mean and maximum surface field strengths, the fraction of the total magnetic energy stored in the poloidal component, the percentages of poloidal and toroidal energy stored in dipolar ($l=1$), quadrupolar ($l=2$) and octopolar modes ($l=3$), the percentages of the total, poloidal and toroidal field energies stored in axisymmetric modes and the strength and location of the negative pole of the radial component of the large-scale dipole. The upper and lower values shown for each measured parameter represent their sensitivity to variations in the $\rm{P_{rot}}$ ($^{+0.012}_{-0.011}\rm{\,d}$), $\rm{d\Omega}$ ($\pm0.114\rm{\,rad\,d}^{-1}$), $v\sin i$ ($\pm0.5$\,km\,s$^{-1}$) and RV ($\pm0.1$\,km\,s$^{-1}$). * The 2005 May data originally published in \citet{Jeffers2008} was not reanalysed here. The magnetic field parameters reported in Table \ref{tab:results} were provided by the authors of \citet{Jeffers2008}.}
\setlength{\tabcolsep}{1.3pt}
\begin{tabular}{lccccccccccccccccccc}
\toprule
\multirow{2}{*}{\textbf{Map}} & 
{\textbf{RV}} &
{\textbf{Mean spot}}
& {\textbf{Mean spot}}& {\textbf{Mean bright}}& {\textbf{$\rm{B_{mean}}$}} & {\textbf{$\rm{B_{max}}$}} & \multicolumn{4}{c}{\textbf{Poloidal}}              & \multicolumn{3}{c}{\textbf{Toroidal}}                        & \multicolumn{3}{c}{\textbf{Axisymmetric}}                             & \multicolumn{2}{c}{\textbf{Dipole radial -ve}}              \\ \cmidrule(lr){8-11} \cmidrule(lr){12-14} \cmidrule(lr){15-17} \cmidrule(lr){18-19}
                                   & {\textbf{($kms^{-1}$)}} & {\textbf{(\%, spot-only)}} &  {\textbf{(\%, S+B)}} & {\textbf{(\%, S+B)}}
                                   &  {\textbf{(G)}} & {\textbf{(G)}}                    & \textbf{\% tot} & \textbf{$l=1$} & \textbf{$l=2$} & \textbf{$l=3$} & \textbf{$l=1$} & \textbf{$l=2$} & \textbf{$l=3$} & \textbf{\% tot} & \textbf{\% Pol} & \textbf{\% Tor} & \textbf{B (G)} & \textbf{Lat.} \\ \midrule
2004 Sept.& -22.8$\pm$0.1 &  $6.8^{+1.7}_{-0.6}$  &  $10.5^{+2.6}_{-0.7}$  &  $97.2^{+0.1}_{-0.6}$  &  $95^{+7}_{-3}$  &  $549^{+16}_{-8}$  &  $62^{+2}_{-1}$  &  $37^{+0}_{-1}$  &  $8^{+0}_{-0}$  &  $9^{+1}_{-1}$  &  $2^{+1}_{-0}$  &  $8^{+0}_{-1}$  &  $33^{+1}_{-1}$  &  $52^{+1}_{-2}$  &  $38^{+1}_{-2}$  &  $75^{+1}_{-2}$  &  $-97^{+4}_{-5}$  &  $70^{+0}_{-0}$   \\ [+1.5mm] 
2005 May* & -22.9$\pm$0.1 & - & -  & - & 113 & 462 & 47 & 45 & 3&  10 &1 &2&9 &47 &36& 57& -105& 49 \\ [+1.5mm] 
2007 May& -22.8$\pm$0.1 &  $5.4^{+2.9}_{-0.0}$  &  $8.7^{+4.6}_{-0.0}$  &  $97.4^{+0.1}_{-0.1}$  &  $141^{+22}_{-0}$  &  $912^{+120}_{-0}$  &  $34^{+3}_{-1}$  &  $20^{+0}_{-8}$  &  $17^{+6}_{-0}$  &  $5^{+1}_{-0}$  &  $8^{+1}_{-0}$  &  $6^{+0}_{-1}$  &  $15^{+2}_{-1}$  &  $66^{+0}_{-2}$  &  $35^{+1}_{-4}$  &  $81^{+0}_{-1}$  &  $-84^{+14}_{-0}$  &  $76^{+0}_{-6}$   \\ [+1.5mm] 
2007 Nov.& -22.8$\pm$0.1 &  $5.5^{+2.2}_{-0.4}$  &  $9.2^{+5.5}_{-0.0}$  &  $97.7^{+0.1}_{-0.0}$  &  $136^{+29}_{-5}$  &  $813^{+248}_{-0}$  &  $45^{+0}_{-1}$  &  $24^{+2}_{-6}$  &  $8^{+1}_{-1}$  &  $11^{+1}_{-0}$  &  $2^{+0}_{-0}$  &  $6^{+0}_{-0}$  &  $23^{+1}_{-1}$  &  $40^{+1}_{-0}$  &  $13^{+2}_{-0}$  &  $63^{+1}_{-1}$  &  $-86^{+0}_{-1}$  &  $50^{+0}_{-3}$   \\ [+1.5mm] 
2008 May& -22.9$\pm$0.1 &  $3.6^{+0.9}_{-0.0}$  &  $5.8^{+1.1}_{-0.0}$  &  $98.5^{+0.0}_{-0.2}$  &  $78^{+2}_{-0}$  &  $329^{+0}_{-10}$  &  $85^{+0}_{-1}$  &  $62^{+0}_{-2}$  &  $5^{+1}_{-0}$  &  $4^{+0}_{-0}$  &  $1^{+0}_{-0}$  &  $1^{+0}_{-0}$  &  $1^{+0}_{-0}$  &  $58^{+0}_{-2}$  &  $66^{+0}_{-2}$  &  $15^{+2}_{-1}$  &  $-110^{+1}_{-1}$  &  $70^{+0}_{-1}$   \\ [+1.5mm] 
2009 May& -22.9$\pm$0.1 &  $7.2^{+2.3}_{-0.0}$  &  $8.8^{+2.3}_{-0.1}$  &  $97.8^{+0.0}_{-0.5}$  &  $123^{+15}_{-8}$  &  $643^{+102}_{-15}$  &  $52^{+1}_{-0}$  &  $28^{+0}_{-2}$  &  $8^{+1}_{-0}$  &  $9^{+0}_{-0}$  &  $2^{+1}_{-0}$  &  $3^{+0}_{-0}$  &  $19^{+1}_{-1}$  &  $53^{+0}_{-1}$  &  $36^{+0}_{-1}$  &  $72^{+0}_{-1}$  &  $-107^{+6}_{-8}$  &  $90^{+0}_{-1}$   \\ [+1.5mm] 
2010 Aug.& -22.3$\pm$0.2 & $6.9^{+2.8}_{-0.0}$  &  $10.5^{+2.2}_{-0.0}$  &  $97.3^{+0.0}_{-0.5}$  &  $156^{+20}_{-5}$  &  $1239^{+71}_{-87}$  &  $47^{+3}_{-0}$  &  $9^{+3}_{-2}$  &  $12^{+1}_{-2}$  &  $15^{+0}_{-1}$  &  $10^{+2}_{-0}$  &  $3^{+1}_{-1}$  &  $18^{+1}_{-2}$  &  $54^{+0}_{-4}$  &  $32^{+2}_{-6}$  &  $73^{+1}_{-2}$  &  $-47^{+2}_{-17}$  &  $28^{+16}_{-14}$    \\ [+1.5mm] 
2011 May 14-16& -22.7$\pm$0.2 &$6.9^{+1.3}_{-0.0}$  &  $12.1^{+1.9}_{-0.0}$  &  $96.6^{+0.0}_{-0.5}$  &  $158^{+16}_{-4}$  &  $938^{+78}_{-52}$  &  $64^{+1}_{-1}$  &  $31^{+0}_{-4}$  &  $25^{+4}_{-1}$  &  $10^{+1}_{-1}$  &  $6^{+0}_{-0}$  &  $9^{+0}_{-1}$  &  $21^{+1}_{-1}$  &  $60^{+1}_{-2}$  &  $59^{+1}_{-3}$  &  $64^{+1}_{-0}$  &  $-131^{+4}_{-0}$  &  $58^{+0}_{-4}$   \\ [+1.5mm] 
2011 May 17-19& -22.6$\pm$0.2 &  $7.1^{+2.8}_{-0.0}$  &  $11.6^{+2.4}_{-0.0}$  &  $96.7^{+0.0}_{-0.6}$  &  $158^{+22}_{-6}$  &  $840^{+106}_{-54}$  &  $69^{+2}_{-0}$  &  $28^{+0}_{-3}$  &  $22^{+5}_{-0}$  &  $11^{+0}_{-1}$  &  $9^{+0}_{-1}$  &  $6^{+0}_{-1}$  &  $19^{+1}_{-0}$  &  $54^{+2}_{-3}$  &  $51^{+3}_{-4}$  &  $62^{+3}_{-3}$  &  $-116^{+12}_{-5}$  &  $58^{+0}_{-6}$   \\ [+1.5mm] 
2013 Sept. 11-15& -22.6$\pm$0.1 &  $7.0^{+1.8}_{-0.1}$  &  $9.3^{+3.2}_{-0.7}$  &  $97.9^{+0.2}_{-0.6}$  &  $102^{+1}_{-0}$  &  $530^{+0}_{-27}$  &  $69^{+2}_{-1}$  &  $25^{+1}_{-0}$  &  $14^{+1}_{-2}$  &  $12^{+2}_{-2}$  &  $31^{+3}_{-2}$  &  $22^{+1}_{-1}$  &  $16^{+2}_{-3}$  &  $22^{+2}_{-3}$  &  $5^{+1}_{-1}$  &  $60^{+2}_{-4}$  &  $-42^{+3}_{-4}$  &  $24^{+7}_{-3}$    \\ [+1.5mm] 
2018 June& -22.9$\pm$0.1&  $5.4^{+1.3}_{-0.3}$  &  $8.5^{+1.9}_{-1.1}$  &  $98.0^{+0.3}_{-0.5}$  &  $87^{+4}_{-3}$  &  $357^{+10}_{-0}$  &  $51^{+7}_{-0}$  &  $26^{+4}_{-6}$  &  $5^{+2}_{-1}$  &  $17^{+3}_{-3}$  &  $4^{+2}_{-1}$  &  $4^{+0}_{-1}$  &  $44^{+0}_{-3}$  &  $54^{+1}_{-9}$  &  $24^{+3}_{-6}$  &  $84^{+0}_{-3}$  &  $-65^{+2}_{-2}$  &  $68^{+0}_{-3}$   \\ [+1.5mm] 
2018 July 17-28&-22.8$\pm$0.1 &  $5.7^{+2.2}_{-0.6}$  &  $9.1^{+2.1}_{-0.9}$  &  $97.6^{+0.2}_{-0.6}$  &  $134^{+15}_{-0}$  &  $703^{+0}_{-70}$  &  $47^{+4}_{-1}$  &  $22^{+2}_{-3}$  &  $6^{+0}_{-1}$  &  $13^{+5}_{-1}$  &  $4^{+2}_{-1}$  &  $4^{+0}_{-0}$  &  $32^{+2}_{-4}$  &  $46^{+0}_{-5}$  &  $10^{+1}_{-1}$  &  $78^{+0}_{-5}$  &  $-78^{+0}_{-8}$  &  $42^{+1}_{-4}$   \\ [+1.5mm] 
2019 June&-22.8$\pm$0.1 &  $5.9^{+1.4}_{-0.5}$  &  $9.7^{+2.1}_{-0.9}$  &  $97.5^{+0.2}_{-0.6}$  &  $149^{+30}_{-8}$  &  $1152^{+268}_{-81}$  &  $60^{+0}_{-2}$  &  $11^{+2}_{-0}$  &  $17^{+0}_{-1}$  &  $24^{+0}_{-2}$  &  $4^{+2}_{-1}$  &  $4^{+0}_{-0}$  &  $29^{+0}_{-1}$  &  $41^{+4}_{-5}$  &  $30^{+7}_{-8}$  &  $57^{+0}_{-2}$  &  $-69^{+0}_{-15}$  &  $-12^{+9}_{-22}$   \\ [+1.5mm] 
2019 July 10-19&-22.8$\pm$0.1 &  $6.0^{+1.2}_{-0.6}$  &  $8.8^{+2.1}_{-0.8}$  &  $98.1^{+0.1}_{-0.5}$  &  $124^{+12}_{-5}$  &  $438^{+63}_{-0}$  &  $69^{+4}_{-5}$  &  $27^{+1}_{-5}$  &  $14^{+0}_{-3}$  &  $11^{+1}_{-2}$  &  $9^{+2}_{-2}$  &  $4^{+3}_{-2}$  &  $29^{+2}_{-6}$  &  $45^{+1}_{-2}$  &  $36^{+2}_{-6}$  &  $64^{+9}_{-6}$  &  $-109^{+14}_{-10}$  &  $74^{+6}_{-6}$   \\ [+1.5mm] 
2019 July 20-29&-22.8$\pm$0.1 &  $5.5^{+1.2}_{-0.4}$  &  $9.4^{+1.9}_{-0.9}$  &  $97.5^{+0.2}_{-0.4}$  &  $165^{+32}_{-7}$  &  $1082^{+258}_{-19}$  &  $45^{+2}_{-1}$  &  $22^{+3}_{-5}$  &  $7^{+1}_{-0}$  &  $10^{+1}_{-0}$  &  $1^{+0}_{-0}$  &  $2^{+0}_{-0}$  &  $32^{+2}_{-3}$  &  $50^{+0}_{-2}$  &  $21^{+1}_{-1}$  &  $74^{+0}_{-2}$  &  $-115^{+1}_{-9}$  &  $62^{+0}_{-1}$   \\ [+1.5mm]

\bottomrule
    \end{tabular}
    \label{tab:results}
\end{table*}

\begin{figure*}
    \centering
    \includegraphics[width=\linewidth]{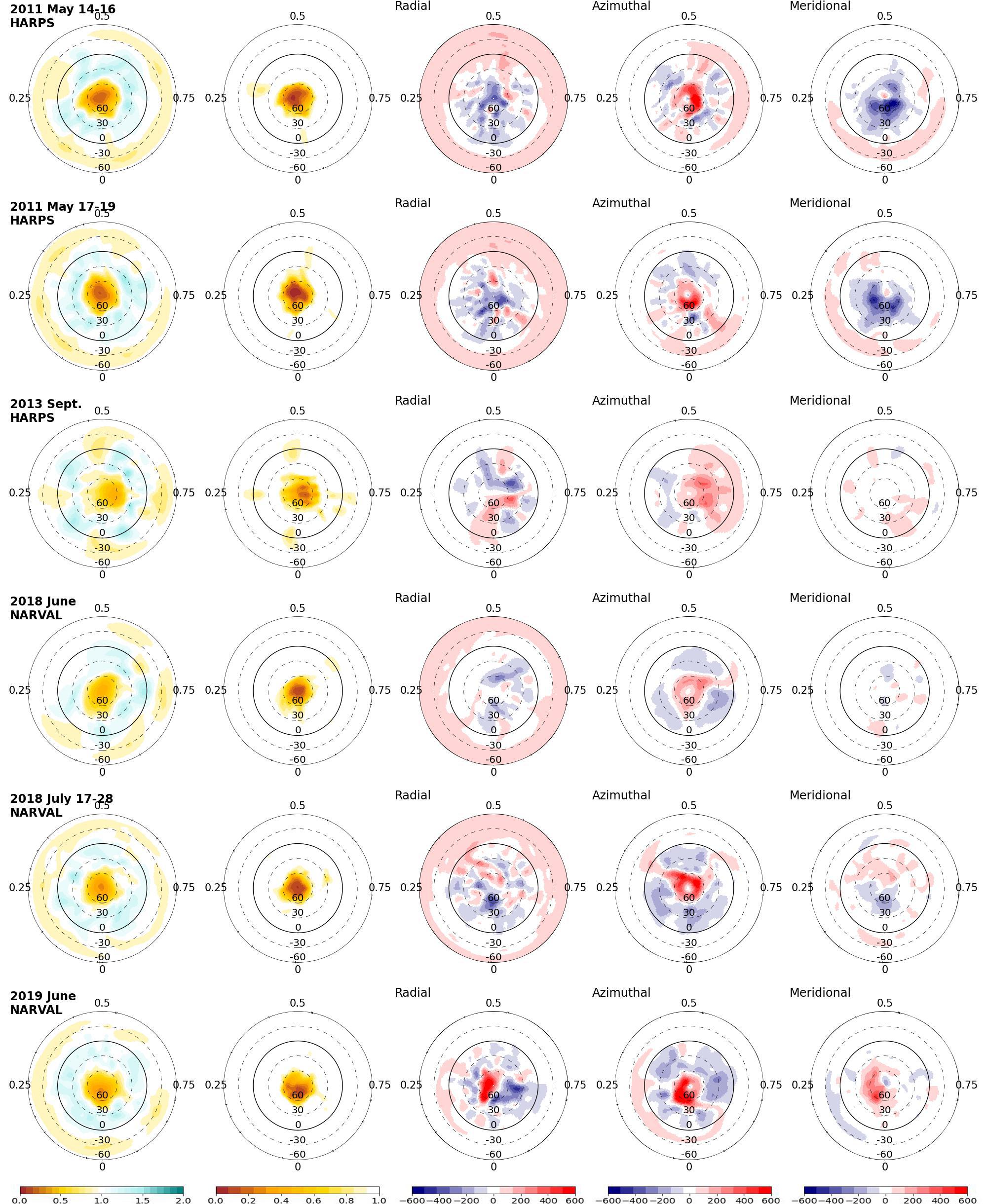}
    \caption{Same as Figure \ref{fig:maps1} but for 2011 May 14-16 through to 2019 June.}
    \label{fig:maps2}
\end{figure*}

\begin{figure*}
    \centering
    \includegraphics[width=\linewidth]{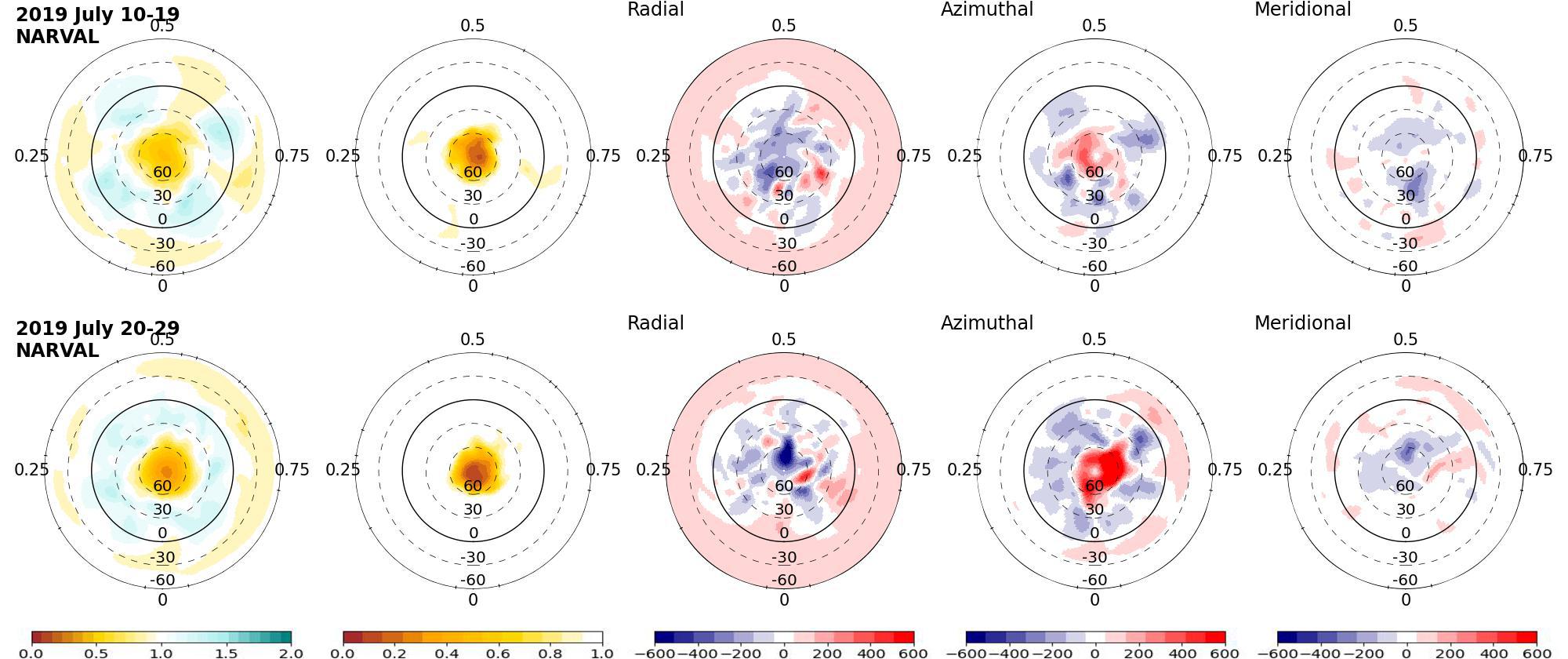}
    \caption{Same as Figure \ref{fig:maps1} but for 2019 July 10-19 and 2019 July 20-29.}
    \label{fig:maps3}
\end{figure*}

\section{Impact of brightness topology as an input to magnetic field modelling}

\begin{figure*}
    \centering
    \includegraphics[width=0.65\linewidth]{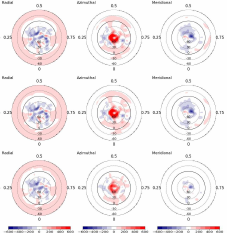}
    \caption{Top: The magnetic field in 2007 May when the brightness + spot model is considered during magnetic field reconstruction. Middle: The magnetic field as shown in Figure \ref{fig:maps1} when the spot-only (filling factor) model is considered during magnetic field reconstruction. Bottom: The magnetic field when no brightness model is considered in the reconstruction of the magnetic model. The azimuthal and meridional fields are clearly weaker when the presence of the dark polar spot is not considered in the magnetic field reconstruction. There is minimal difference in the radial magnetic field reconstruction across the three models. There is also minimal difference in the three reconstructed field components when using the brightness + spot or the spot-only models as input for the field reconstruction.  }
    \label{fig:Appendix_may01maps}
\end{figure*}

Figure \ref{fig:Appendix_may01maps} shows the impacts on the final reconstructed magnetic field of using pre-computed DI maps for the magnetic field inversion. 

\section{Results of RV activity filtering - GPR method}\label{sec:GP_results_full}

\begin{figure*}
    \centering
    \includegraphics[width=0.49\linewidth]{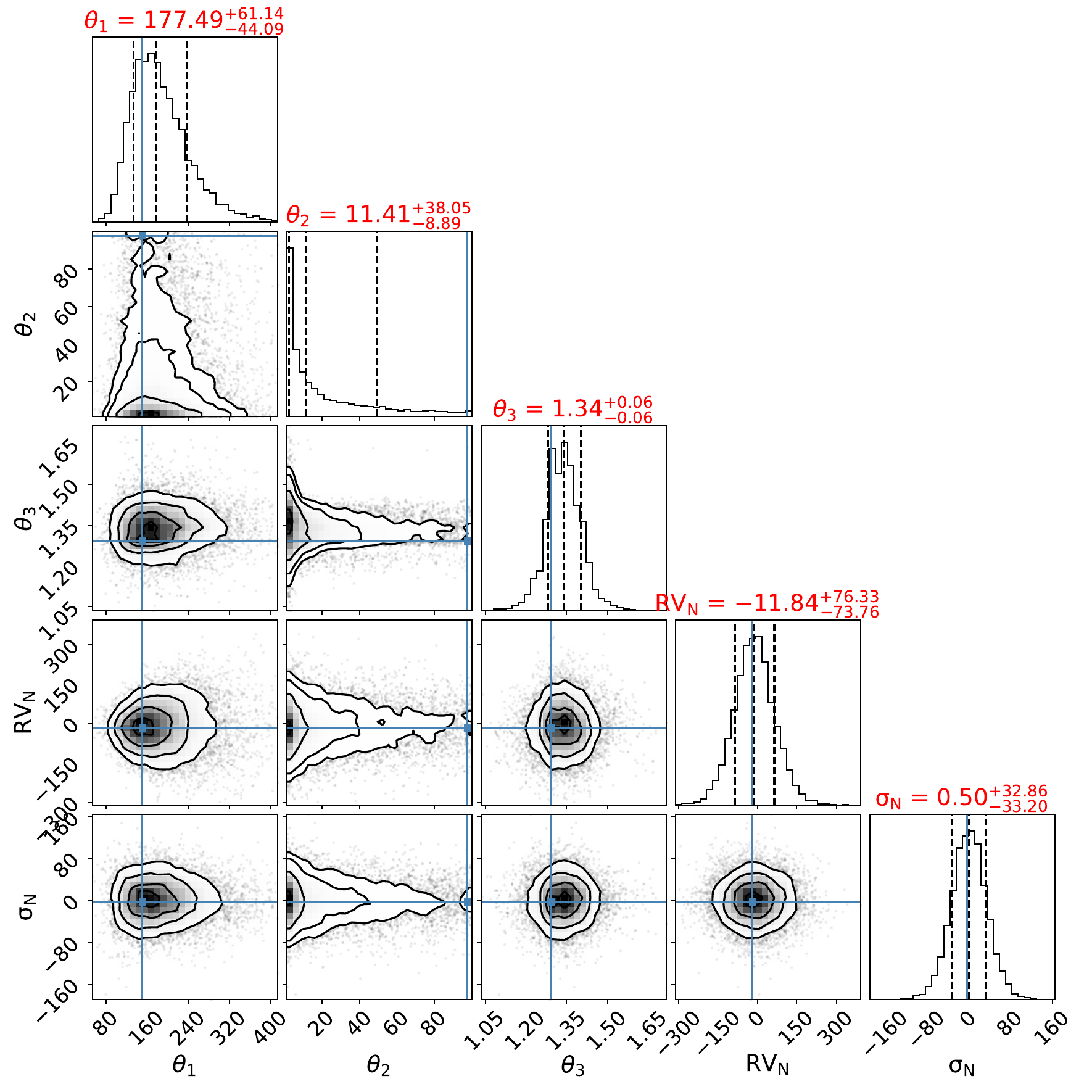}
    \includegraphics[width=0.49\linewidth]{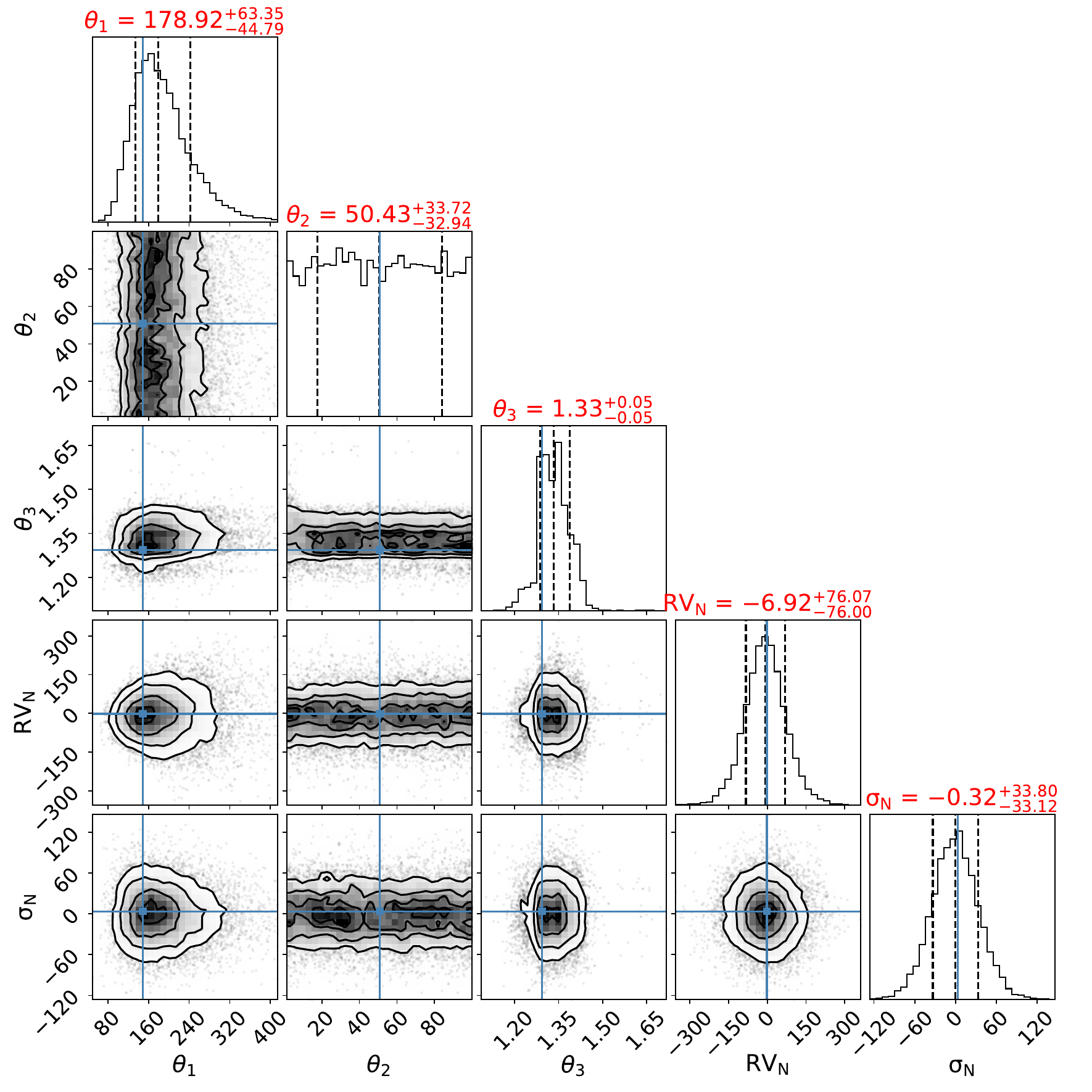}
    \caption{Posterior distribution of the fitted parameters for an activity-only model to the RVs for 2007 May. Left: the priors for each parameter are as per Table \ref{tab:GP_priors}. Right: the priors for $\theta_2$ are $\mathcal{U}$ (1.33, 100). The blue lines indicate the most likely set of parameters for the GP and dashed vertical lines in the histograms are the 0.16, 0.50, and 0.84 quantiles. Contours are 1-$\sigma$, 2-$\sigma$, and 3-$\sigma$ levels. }
    \label{fig:2007may_GPresults}
\end{figure*}

\begin{figure*}
    \centering
    \includegraphics[width=0.49\linewidth]{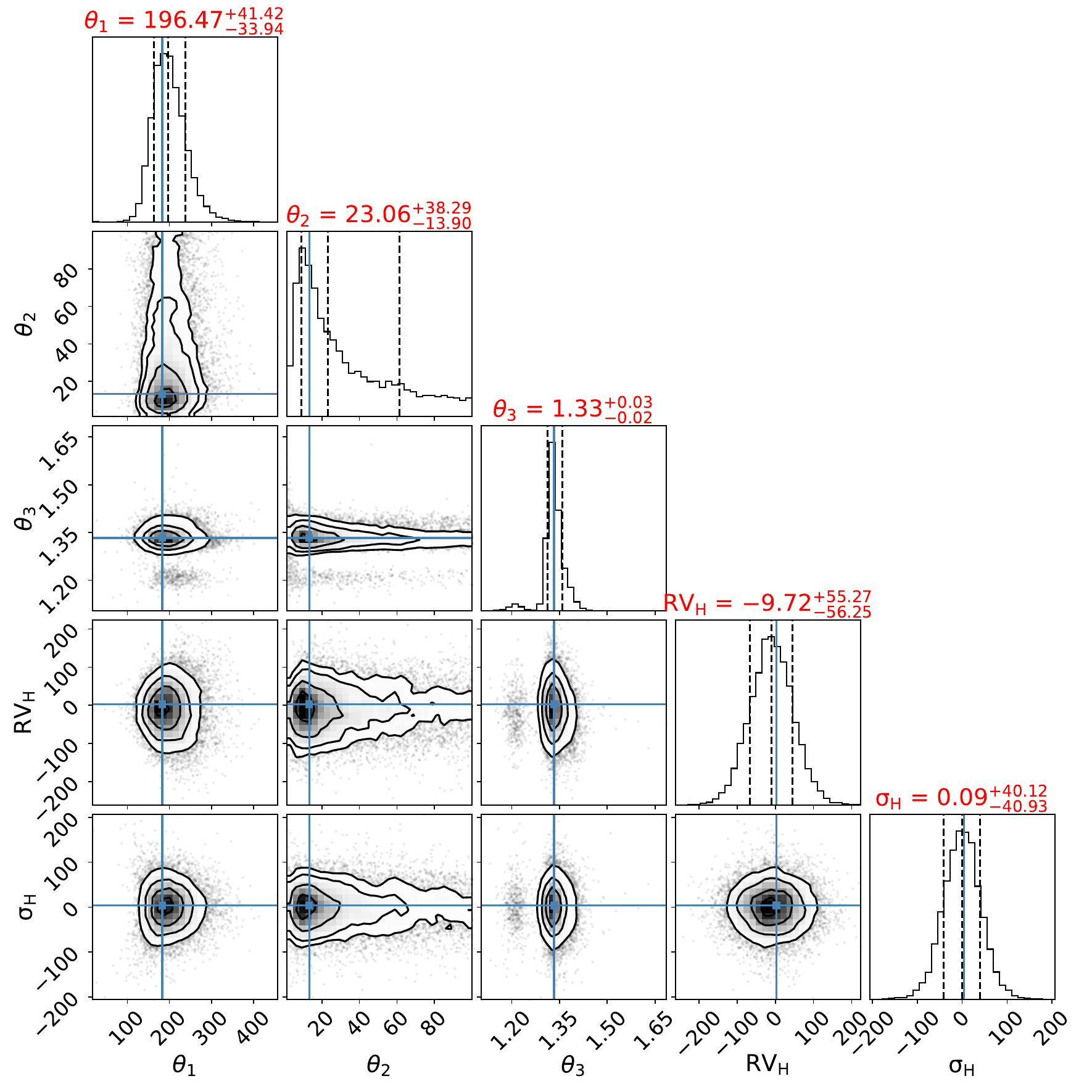}
    \includegraphics[width=0.49\linewidth]{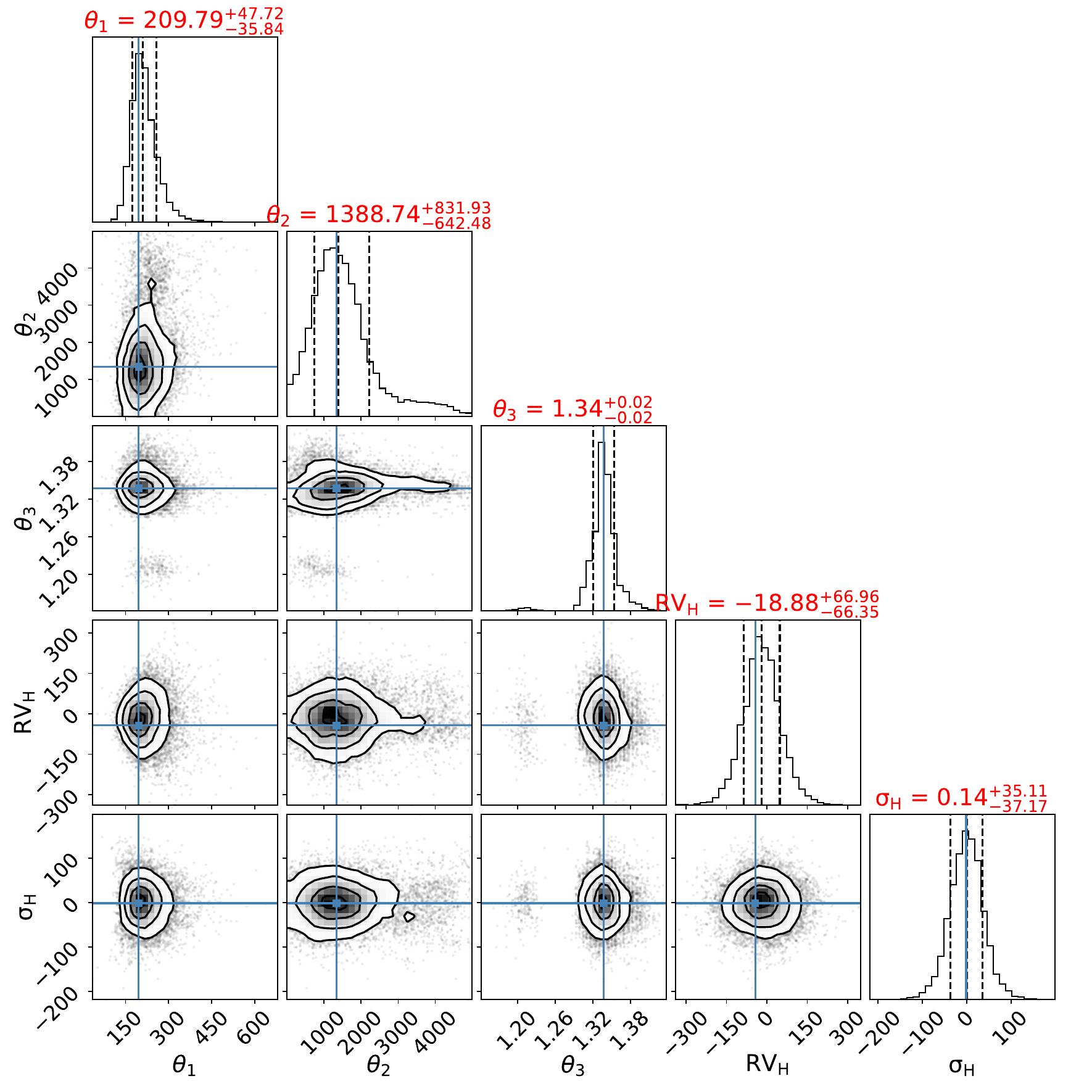}
    \caption{Posterior distribution of the fitted parameters for an activity-only model to the RVs for HARPS data (2010 Aug. to 2013 Sept.). Left:  the priors for each parameter are as per Table \ref{tab:GP_priors}, and the marginal likelihood for this model is $ln Z= -226.5$. Right: the priors for $\theta_2$ are $\mathcal{U}$ (1.33, 5000) and the marginal likelihood for this model is $ln Z= -225.2$.}
    \label{fig:HARPS_GPresults}
\end{figure*}

\begin{figure*}
    \centering
    \includegraphics[width=\linewidth]{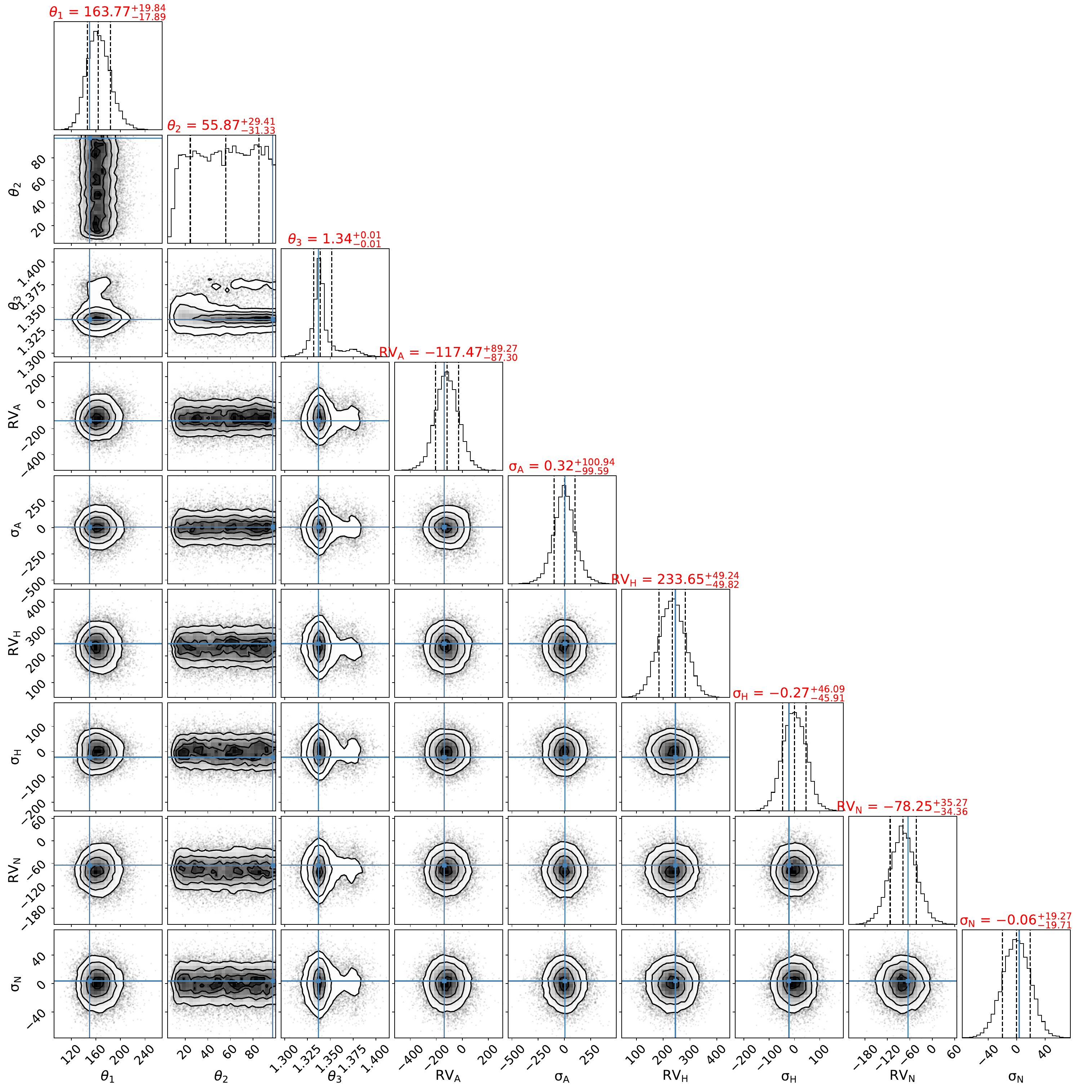}
    \caption{Posterior distribution of the fitted parameters for an activity-only model to the entire set of RVs, from 2004 Sept. through to 2019 July 20-29.}
    \label{fig:AllData_GPresults}
\end{figure*}

Figures \ref{fig:2007may_GPresults} to \ref{fig:AllData_GPresults} show corner plots indicating the posterior distributions and best-fitting parameters for a quasi-periodic GP fitted to various sets of RV observations. 

Figure \ref{fig:2007may_GPresults} (left) shows the corner plot for 2007 May when using a logarithmic prior for $\theta_2$. The logarithmic prior allows for greater exploration of short length scales. In practice this resulted in the decay parameter converging toward the lower edge of the prior (i.e. toward zero, see Figure \ref{fig:2007may_GPresults}) such that the final numerical model did not represent a realistic physical solution. With a uniform prior (right), $\theta_2$ did not converge to a unique solution, regardless of the upper limit of the prior.  Without any strong evidence for the mean or standard deviation of the spot decay timescale, we avoided using an informative Gaussian prior for $\theta_2$, which would bias the model and artificially boost the posterior likelihood. 

Figure \ref{fig:HARPS_GPresults} shows the posterior distributions of the model parameters for the combined HARPS data using two different priors for $\theta_2$. On the left the GP converges toward a short decay timescale of $\sim9$\,d, which may be related to the lengths of individual observing runs. On the right the value for $\theta_2$ is larger than the total length of the data ($\sim1130$\,d), and is possibly tied to the long-lived polar spot which is present for the entire data set.

%\begin{figure*}
  %  \centering
  %  \includegraphics[width=0.49\linewidth]{Figures/may07_0p_GP_corner_plot.pdf}
  %  \caption{Same as Figure \ref{fig:2007may_GPresults}, but the priors for $\theta_2$ are $\mathcal{U}$ (1.33, 100).}
  %  \label{fig:2007may_GPresults2}
%\end{figure*}

%\begin{figure*}
 %   \centering
 %   \includegraphics[width=0.49\linewidth]{Figures/Harps_longDecay_0p_GP_corner_plot.pdf}
 %   \caption{Same as Figure \ref{fig:HARPS_GPresults}, but the priors for $\theta_2$ are $\mathcal{U}$ (1.33, 5000). The marginal likelihood for this model is $ln Z= -223.9$.}
%    \label{fig:HARPS_GPresults2}
%\end{figure*}

%%%%%%%%%%%%%%%%%%%%%%%%%%%%%%%%%%%%%%%%%%%%%%%%%%

%%%%%%%%%%%%%%%%% APPENDICES %%%%%%%%%%%%%%%%%%%%%

\appendix

%%%%%%%%%%%%%%%%%%%%%%%%%%%%%%%%%%%%%%%%%%%%%%%%%%

% Don't change these lines
\bsp	% typesetting comment
\label{lastpage}
\end{document}